\documentclass[sigconf]{acmart}
\copyrightyear{2026}
\acmYear{2026}
\setcopyright{cc}
\setcctype{by-nc-nd}
\acmConference[WWW '26] {Proceedings of the ACM Web Conference 2026}{April 13--17, 2026}{Dubai, United Arab Emirates.}
\acmBooktitle{Proceedings of the ACM Web Conference 2026 (WWW '26), April 13--17, 2026, Dubai, United Arab Emirates}
\acmISBN{979-8-4007-2307-0/2026/04}
\acmDOI{10.1145/XXXXXX.XXXXXX}

\usepackage{amsmath,amsfonts}
\usepackage{graphicx}
\usepackage{textcomp}
\usepackage{xcolor}
\def\BibTeX{{\rm B\kern-.05em{\sc i\kern-.025em b}\kern-.08em
    T\kern-.1667em\lower.7ex\hbox{E}\kern-.125emX}}

 \usepackage{tgtermes}
\usepackage{array}
\usepackage[caption=false,font=footnotesize,labelfont=rm,textfont=rm]{subfig}
\usepackage{caption}
\usepackage{textcomp}
\usepackage{stfloats}
\usepackage{url}
\usepackage{verbatim}
\usepackage{graphicx}
\usepackage{gensymb}
\usepackage{multirow}
\usepackage{booktabs}
\usepackage{colortbl}
\usepackage{makecell}
\usepackage{enumitem}
\usepackage{xcolor}
\usepackage{tikz}
\usepackage{amsmath}
\usepackage[linesnumbered,ruled,vlined,algo2e]{algorithm2e}
\usepackage[ruled, vlined, linesnumbered]{algorithm2e}
\usepackage{algorithm}
\usepackage{algpseudocode}
\usepackage{multirow}
\usepackage{color}
\usepackage{xspace}

\newcommand\SystemName{\textsc{SuperEar}\xspace}

\newcommand\cparagraph[1]{\vspace{1.2mm}\noindent \textbf{#1.}}

\begin{document}
\def\allfiles{}
\date{}

\title{\SystemName: Eavesdropping on Mobile Voice Calls via Stealthy Acoustic Metamaterials}

\author{Zhiyuan Ning}
\email{ningzhiyuan@stumail.nwu.edu.cn}
\orcid{0009-0009-4416-2080}
\affiliation{%
\institution{ACSS Lab, NorthWest University}
  \city{Xi’an}
  \state{Shaanxi}
  \country{China}
}

\author{Zhanyong Tang}
\authornote{Corresponding author\\
PINE Lab: Shaanxi Key Laboratory of Passive Internet of Things and Neural Computing\\
ACSS Lab: Xi’an Key Laboratory of Advanced Computing and Software Security}
\email{zytang@nwu.edu.cn}
\orcid{0000-0002-4333-2334}
\affiliation{%
\institution{ACSS Lab, NorthWest University}
  \city{Xi’an}
  \state{Shaanxi}
  \country{China}
}

\author{Juan He}
\email{hejuan@stumail.nwu.edu.cn}
\orcid{0000-0002-4333-2334}
\affiliation{%
\institution{PINE Lab, NorthWest University}
  \city{Xi’an}
  \state{Shaanxi}
  \country{China}
}

\author{Weizhi	Meng}
\email{weizhi.meng@ieee.org}
\orcid{0000-0003-4384-5786}
\affiliation{%
\institution{Lancaster University}
\city{Lancaster}
\state{Lancashire}
  \country{United Kingdom}
}

\author{Yuntian	Chen}
\email{chenyt@stumail.nwu.edu.cn}
\orcid{0000-0002-2320-9582}
\affiliation{%
\institution{PINE Lab, NorthWest University}
  \city{Xi’an}
  \state{Shaanxi}
  \country{China}
}

\author{Jie	Zhang}
\email{jiezhang@xupt.edu.cn}
\orcid{0000-0001-9176-170X}
\affiliation{%
\institution{Xi'an University of Posts and Telecommunications}
  \city{Xi’an}
  \state{Shaanxi}
  \country{China}
}

\author{Zheng Wang}
\email{z.wang5@leeds.ac.uk}
\orcid{0000-0001-6157-0662}
\affiliation{%
\institution{University of Leeds}
\city{Leeds}
\state{West Yorkshire}
\country{United Kingdom}
}
\renewcommand{\shortauthors} {Zhiyuan Ning et al.}
\begin{abstract}
Acoustic eavesdropping is a privacy risk, but existing attacks rarely work in real outdoor situations where people make phone calls on the move. We present \SystemName, the first portable system that uses acoustic metamaterials to reliably capture conversations in these scenarios. We show that the threat is real as a practical prototype can be implemented to enhance faint signals, cover the full range of speech with a compact design, and reduce noise and distortion to produce clear audio. We show that \SystemName can be implemented from low-cost 3D-printed parts and off-the-shelf hardware. Experimental results show that \SystemName can recover phone call audio with a success rate of over 80\% at distances of up to 4.6 m - more than twice the range of previous approaches. Our findings highlight a new class of privacy threats enabled by metamaterial technology that requires attention. 
\end{abstract}


\maketitle
\section{INTRODUCTION}
Mobile phones are widely used for private voice communication in public environments. Making a phone call while walking outdoors is common in daily life, and users often assume that conversations can be protected by distance, ambient noise, and the limited acoustic output of mobile devices. These assumptions shape how people behave in public space and how mobile systems are designed, yet they are rarely examined under realistic adversarial conditions.

Acoustic eavesdropping attacks exploit unintended sound emissions to recover sensitive information such as account details or transaction data, enabling fraud or identity theft \cite{accear,AccelEve,AccelWord,mmspy}. In noisy outdoor environments, users are particularly vulnerable because human auditory attention tends to focus on the caller’s voice and ignore subtle cues of being monitored \cite{selective1,selective2,selective3}. As a result, acoustic attacks can remain unnoticed even when users are alert to more conventional forms of surveillance.


Prior acoustic eavesdropping attacks leverage motion-sensor~\cite{AccelEve,AccelWord,HDD,Vibphone}, optical~\cite{lamphone,VibraPhone}, and radio frequency (RF)~\cite{mmecho,mmspy} methods. They work in controlled settings but are ineffective in realistic mobile outdoor scenarios. Motion-sensor attacks typically require compromising the victim's device, which is hard to achieve while the target device is moving. Optical and RF approaches require precise targeting of a vibrating surface, which is infeasible when the victim and attacker are moving \cite{lamphone,VibraPhone,mmecho,mmspy}. Acoustic eavesdropping in a moving outdoor environment is also difficult because earpiece audio is very weak and drops below a usable signal-to-noise ratio (SNR) beyond roughly 2 m \cite{mmeve}.

Recent advances in acoustic metamaterials may change this picture. Acoustic metamaterials are engineered structures designed to manipulate sound in ways not achievable with natural materials ~\cite{r11,r12,r14,r15}. By controlling wave interference through their internal geometry, they can amplify, filter, or redirect acoustic signals, with demonstrated applications in areas such as noise reduction and ultrasound imaging.
As a potential eavesdropping medium, metamaterials can enhance the SNR of faint earpiece emissions and capture airborne speech without the need for precise alignment or access to vibrating surfaces. This creates opportunities for eavesdropping attacks that bypass limitations of existing acoustic eavesdropping methods, extending the threat model to scenarios where interception was previously considered impractical.

\begin{figure}[!t]
\centering
\subfloat[]{
		\includegraphics[scale=0.08]{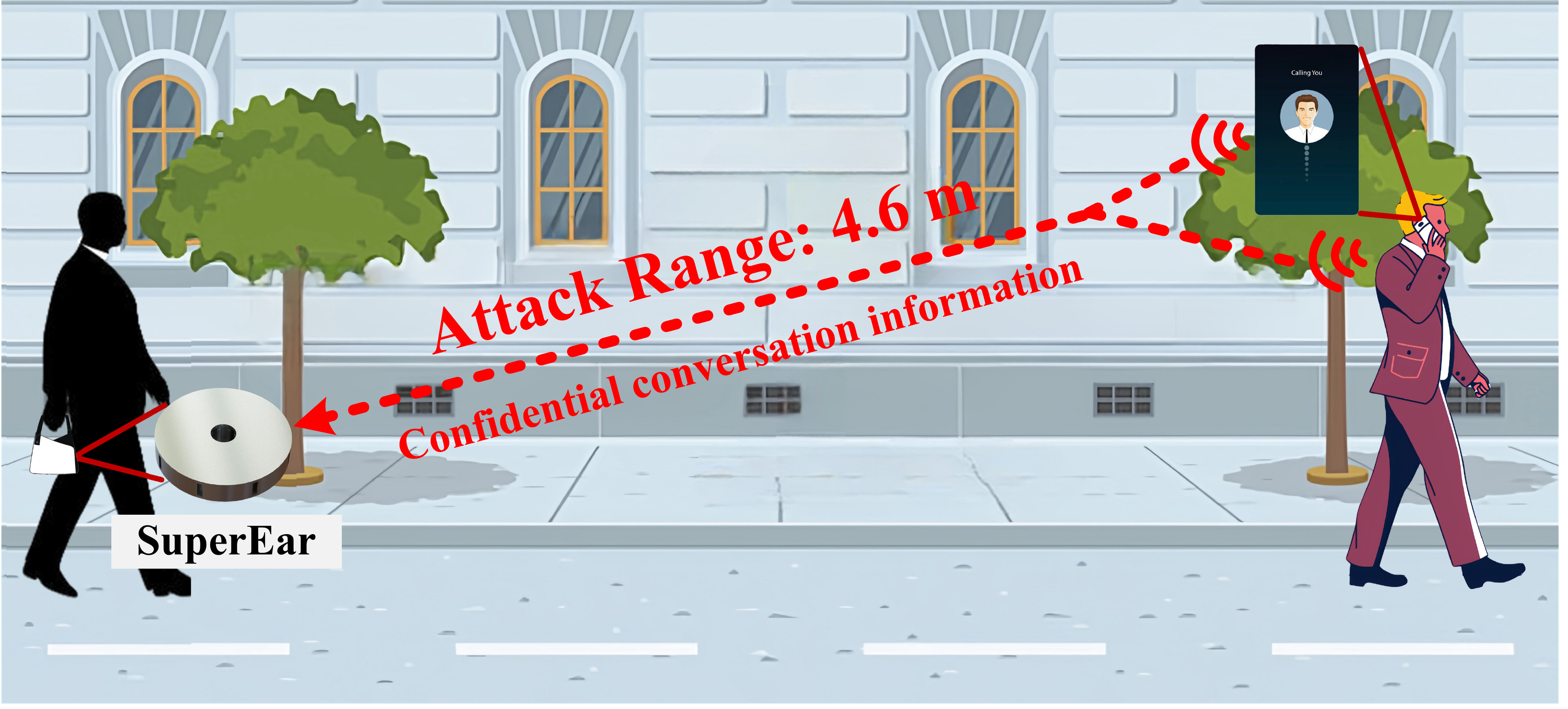}
  \label{A possible attacking scenario}}
\hspace{0.3cm}
\subfloat[]{
		\includegraphics[scale=0.15]{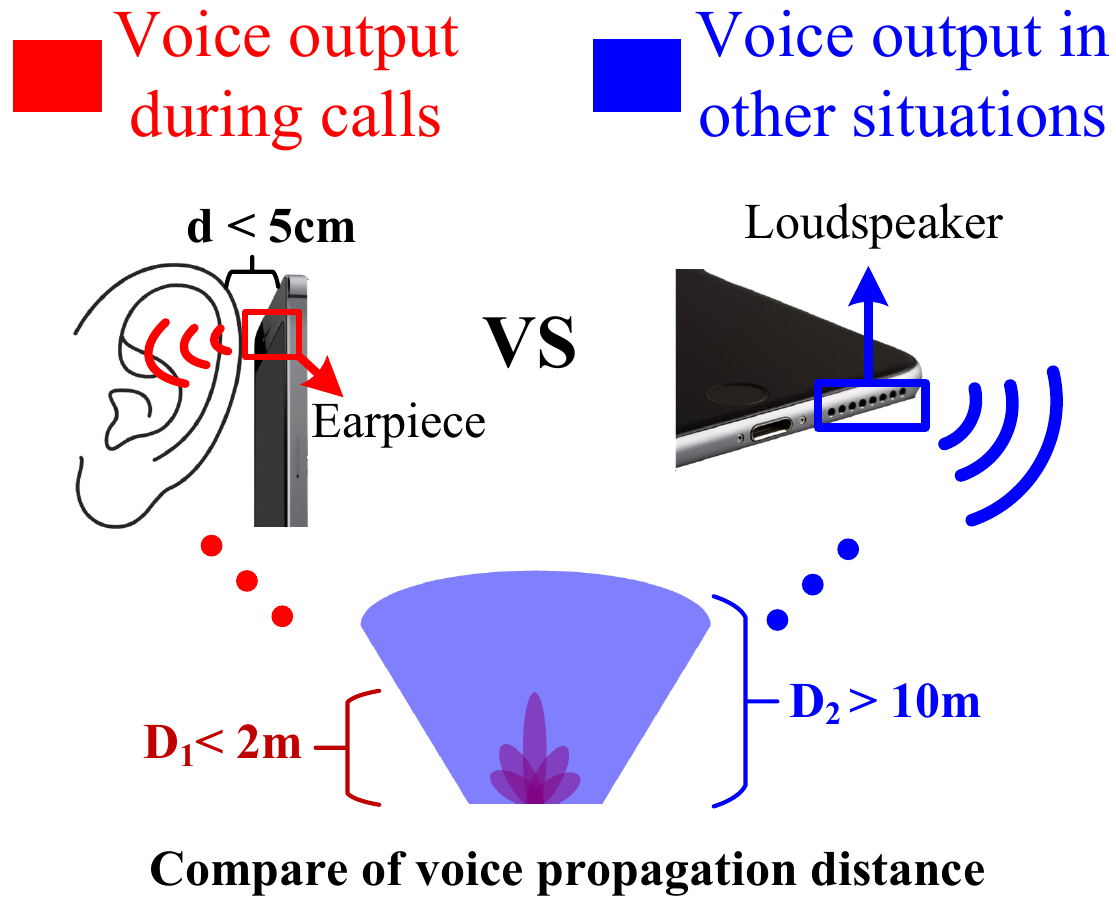}
  \label{earpiece}}
\caption{(a) Possible attack: \SystemName eavesdrops on an outdoor call. (b) Call voice vs. other scenarios.}
\label{12}
\vspace{-2mm}
\end{figure}

This paper demonstrates a new and practical acoustic eavesdropping attack that can intercept phone calls made while a user is walking outdoors, as illustrated in Fig.~\ref{A possible attacking scenario}. We present \SystemName, the first portable system that uses acoustic metamaterials to make outdoor interception of phone calls reliable during everyday travel. 
We show that it is possible to overcome three key limitations that previously prevented metamaterials from being effective for eavesdropping. First, we reduce thermo-viscous losses, i.e., energy lost as heat and friction when sound travels through narrow channels, which otherwise weakens low-frequency signals. Second, we design the system so that the full speech frequency band can be covered with only a few metamaterial elements, keeping the device compact and portable. Finally, we lower the distortion and background noise caused by uneven amplification by applying smoothing and adaptive filtering.  Together, these improvements enable \SystemName to reconstruct clear speech in mobile outdoor conditions.

To show that \SystemName is a real threat, we implemented a working prototype using low-cost, off-the-shelf components, including a Raspberry Pi, a microphone array, and resin 3D-printed acoustic metamaterials. Because the metamaterials are passive, \SystemName requires no additional power source, supporting a compact form factor and long-duration operation. The system captures high-quality audio directly through acoustic design and signal processing, without relying on machine-learning models that can be difficult to build and are hard to adapt to complex environments.

We evaluated \SystemName on nine mobile devices from seven manufacturers under diverse environmental conditions. Results show that \SystemName reliably reconstructs call audio with over 80\% success at distances up to 4.6 m, substantially extending the range of prior attacks and revealing a previously overlooked privacy risk. This paper makes the following contributions.
 
\begin{itemize}[leftmargin=*]
  \item It demonstrates the first practical outdoor acoustic eavesdropping of in‑transit phone calls using acoustic metamaterials;
  \item It identifies and solves three practical challenges for using metamaterials in eavesdropping;
  \item It shows how an attack system can be built from 3D-printed metamaterials and commodity hardware.
\end{itemize}

\section{Background and Related Work} \label{chap:2}




\subsection{Acoustic Eavesdropping Attacks} 
Previous studies on voice communication eavesdropping mainly include RF sensing \cite{mmecho,mmeve,mmspy,WiHear,ART,Tag-Bug,uwhear,waveear,milliear,mmear}, motion sensor–based eavesdropping \cite{AccelEve,AccelWord,accear,PitchIn,speechless,gyrophone,HDD,VibraPhone,V-Speech,earspy}, and optical sensing \cite{VisualMicrophone,lamphone,LidarPhone}. RF sensing reconstructs audio by capturing tiny vibrations with millimeter waves \cite{mmecho,mmspy,mmeve}; motion sensors recover speech by detecting sound-induced micro-vibrations \cite{accear,AccelEve,AccelWord,gyrophone,PitchIn,V-Speech,VibraPhone}; optical sensing remotely captures audio by sensing surface vibrations \cite{lamphone,VisualMicrophone}.

However, RF sensor-based eavesdropping \cite{mmeve,ART,mmspy,mmecho,milliear,Tag-Bug,uwhear,WiHear,waveear,voicelistener} and optical sensor-based eavesdropping \cite{lamphone,VisualMicrophone,LidarPhone} typically rely on fixed observable media, and in outdoor mobile environments, maintaining continuous and precise tracking of the medium becomes challenging, limiting their application. Motion sensor-based eavesdropping \cite{AccelWord,AccelEve,accear,gyrophone,HDD,PitchIn,speechless,Vibphone,V-Speech} requires intrusion into the target device, increasing the risk of exposure for the attacker.

\subsection{Outdoor Eavesdropping on Voice Output} \label{chap:2.2} As shown in Fig.~\ref{earpiece}, during private calls, users typically hold the phone within 5 cm of the ear. Modern earpieces employ beamforming to enhance privacy by directing most of the acoustic energy toward the user’s ear while minimizing leakage into the surrounding air~\cite{mmspy}. As a result, the effective transmission distance of earpiece audio is far shorter than that of the loudspeaker, with the SNR dropping below the detection threshold beyond roughly 2 m~\cite{mmspy,mmeve}. In outdoor scenarios, additional factors such as background noise, wind, and movement further shorten this range, making eavesdropping on phone calls particularly challenging.


\subsection{Acoustic Metamaterials\label{sec:ameta}}
Acoustic metamaterials manipulate sound through engineered internal structures, enabling precise bending, focusing, and amplification for applications such as noise reduction, imaging, and audio enhancement~\cite{focus4,focus5}. Mie resonators, compact space-coiled structures, trap and re-radiate sound at specific frequencies to provide omnidirectional enhancement and can be easily concealed in everyday objects, making them suitable for mobile eavesdropping~\cite{MieResonances,MieResonances2}.

Their limitations include narrowband enhancement, uneven gain, and amplification of ambient noise. This work is the first to show that, despite these constraints, carefully designed acoustic metamaterials can enable practical outdoor eavesdropping on phone calls, revealing a new and realistic privacy threat.




\subsection{Threat Model} \label{chap:2.4} 
Our threat model considers a realistic outdoor acoustic eavesdropping attack on mobile calls, where an adversary captures acoustic leakage from a victim’s phone earpiece during normal walking and conversation. The attacker uses a portable device built from passive acoustic metamaterials, a small microphone array, and commodity hardware, requiring \textbf{no compromise} of the victim’s device or network and relying solely on physical side channels.

To remain covert, the attacker operates beyond a conservative \textbf{safe distance (>3 m)}~\cite{r99,privacy2,privacy3}. The key challenge is recovering intelligible speech from extremely weak earpiece signals in noisy environments. Attacks involving malware, network interception, specialized sensing hardware, or earphone-based calls are outside the scope.

\section{Methodology}
\subsection{Overview of \SystemName}
\label{chap:3}

\begin{figure*}[t]
    \centering
    \includegraphics[width=1\linewidth]{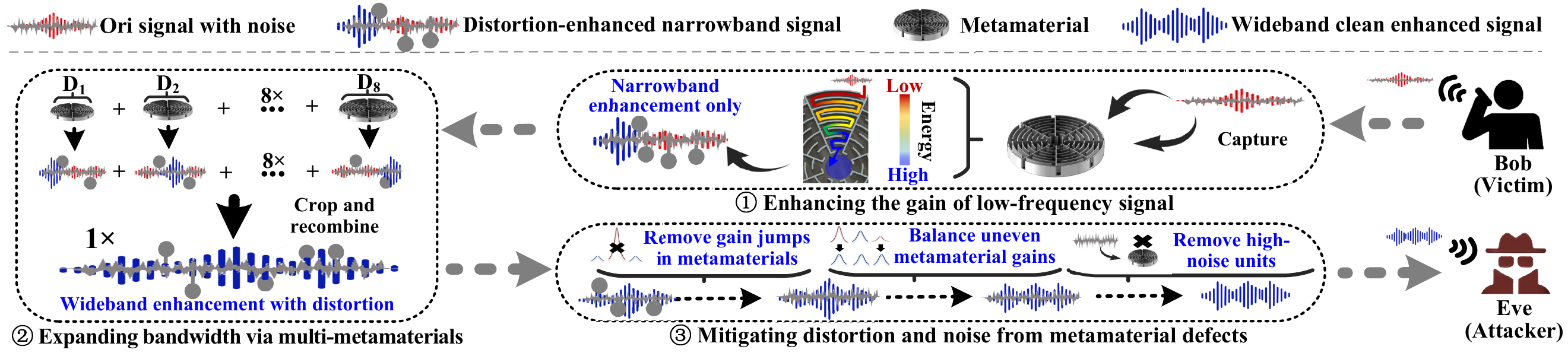}
    \caption{\SystemName eavesdropping overview: enhanced low-frequency narrowband signals (\textbf{Sec.~\ref{chap:6.1}}), wideband amplification via multi-metamaterials (\textbf{Sec.~\ref{chap:6.2}}), and defect-induced distortion/noise suppression (\textbf{Sec.~\ref{chap:6.3}}).}
    \label{Fb overview}
    \vspace{-3mm}
\end{figure*}

We implemented a \SystemName prototype using a Raspberry Pi, eight acoustic metamaterials, and a microphone array (see Sec.~\ref{chap:5}). As shown in Fig.~\ref{Fb overview}, the eavesdropping process consists of three stages: signal capture and enhancement, cropping and recombination, and distortion suppression with noise reduction.

\cparagraph{Enhancing the gain of low-frequency signal}
When the target initiates a call, \SystemName captures acoustic signals using metamaterials and applies the Minimum Variance Distortionless Response (MVDR) algorithm~\cite{MVDR,MVDR2} to extract the earpiece audio while suppressing interference from other directions. Since the target’s position is directly observable, the system remains behind the target and follows along the radial path, allowing MVDR to focus on that direction and improving the stability and accuracy of direction estimation (\textit{\textbf{V1}}, see Sec.~\ref{chap:6.1}).

\cparagraph{Expanding bandwidth via multi-metamaterials}
To obtain a wideband signal, \SystemName builds a multi-metamaterial system and crops and recombines the frequency bands enhanced by each metamaterial, reconstructing a complete voice signal covering all bands (\textit{\textbf{V2}}) (see Sec.~\ref{chap:6.2} for details).

\cparagraph{Mitigating distortion and noise from metamaterial defects}
The system first applies a distortion suppression algorithm to eliminate gain jumps caused by metamaterial defects and balance gains across frequency bands, producing high-fidelity audio (\textit{\textbf{V3}}). It then analyzes the background noise spectrum and uses a noise suppression algorithm with SoX~\cite{Sox} to filter interference, yielding the processed signal (\textit{\textbf{V4}}) (see Sec.~\ref{chap:6.3} for details).

\subsection{Technical Challenges}
There are several obstacles that limit the practicality of acoustic metamaterials for eavesdropping.
The first challenge is improving performance at low frequencies. Speech signals carry important information in this range, but existing designs lose much of their energy as heat when trying to amplify low-frequency sounds.
The second challenge is covering the speech frequency bands while avoiding a large, bulky device. As a single metamaterial only amplifies a narrow frequency band, covering the full speech range typically requires many units.
The third challenge is maintaining audio quality in realistic environments. Manufacturing imperfections can introduce sudden jumps in amplification, while uneven gain across frequencies can distort the reconstructed signal. In addition, metamaterials amplify environmental noise, such as traffic and wind, together with the target speech, further reducing clarity.  

These obstacles have so far limited the use of acoustic metamaterials in realistic attacks. In the following sections, we describe how \SystemName overcomes these issues and makes outdoor eavesdropping feasible.






\subsection{Enhancing the Gain of Low-frequency Signal} \label{chap:6.1}
\begin{figure}[!t]
\vspace{-3mm}
\centering
\subfloat[]{
		\includegraphics[scale=0.14]{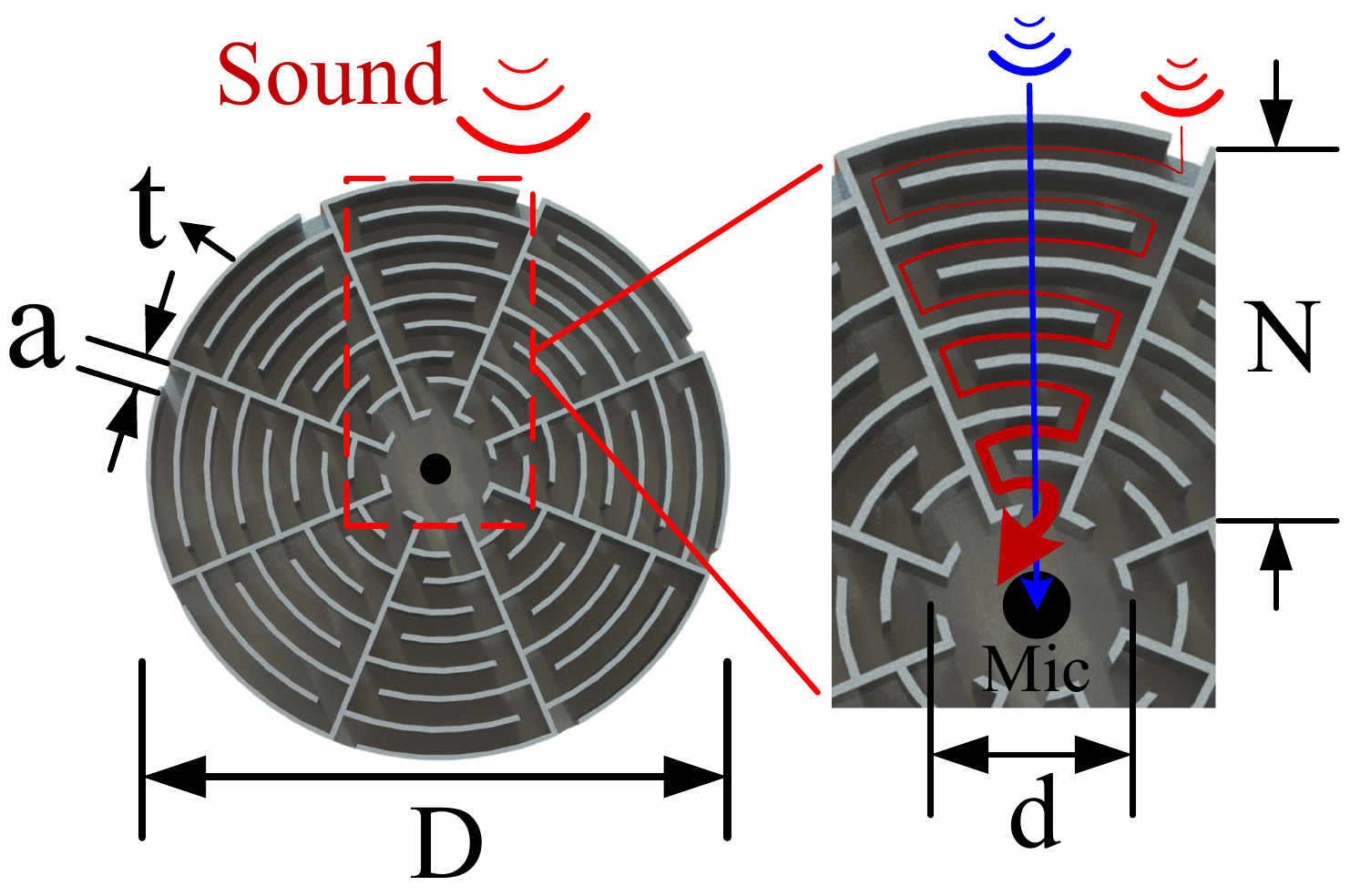}
  \label{structure}}
\hspace{0.3cm}
\subfloat[]{
		\includegraphics[scale=0.42]{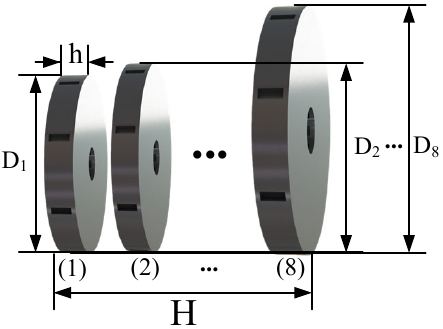}
  \label{structuresys}}
\caption{(a) Acoustic metamaterial structure with a microphone at the center. (b) Metamaterial system.}
\label{LAM-A}
\vspace{-10pt}
\end{figure}

As described in Sec.~\ref{sec:ameta}, we use Mie resonators to prototype \SystemName. We start by analyzing how acoustic metamaterials based on Mie resonators use coiled channels to enhance sound pressure and identified their limitations in low-frequency performance.  

As shown in Fig.~\ref{structure}, the resonator structure consists of a curved propagation path with path width \(a = 3.2 \, \text{mm}\), wall thickness \(t = 0.8 \, \text{mm}\), outer diameter \(D = 80 \, \text{mm}\), inner diameter \(d = 14 \, \text{mm}\), and \(N = 8\) turns. The sound pressure detected at the central microphone can be expressed as~\cite{meta,MieResonances2}:  
\[
P = P_0 \cdot \frac{n_r(f_r)}{\lambda_0} \cdot \sqrt{\frac{2 \rho c^2}{\lambda_0^2}},
\]  
where \(P_0\) is the input sound pressure, \(n_r\) the effective refractive index, \(\lambda_0\) the wavelength in air, \(\rho\) the medium density, and \(c\) the speed of sound. Increasing \(n_r\) concentrates more acoustic energy at the microphone. The gain is strongest near the resonant frequency \(f_r\). For the parameters above, the resonator exhibits a resonance at 563 Hz, enhancing sound in a narrow 100 Hz band with up to 16× gain~\cite{meta}.  

Prior work attempted to shift the resonance into the lower speech range by increasing the number of channels \(N\). For example, setting \(N=22\) lowers \(f_r\) to around 260 Hz~\cite{ultra}. However, longer coiled paths also increase thermo-viscous losses, as sound reflections and scattering dissipate more energy as heat. This limits achievable gain in the 250 - 300 Hz range to about 5×, far below what is needed for reliable eavesdropping.  

\begin{figure}[t!]
\centering
\setlength{\abovecaptionskip}{3pt}
\subfloat[]{
		\includegraphics[scale=0.045]{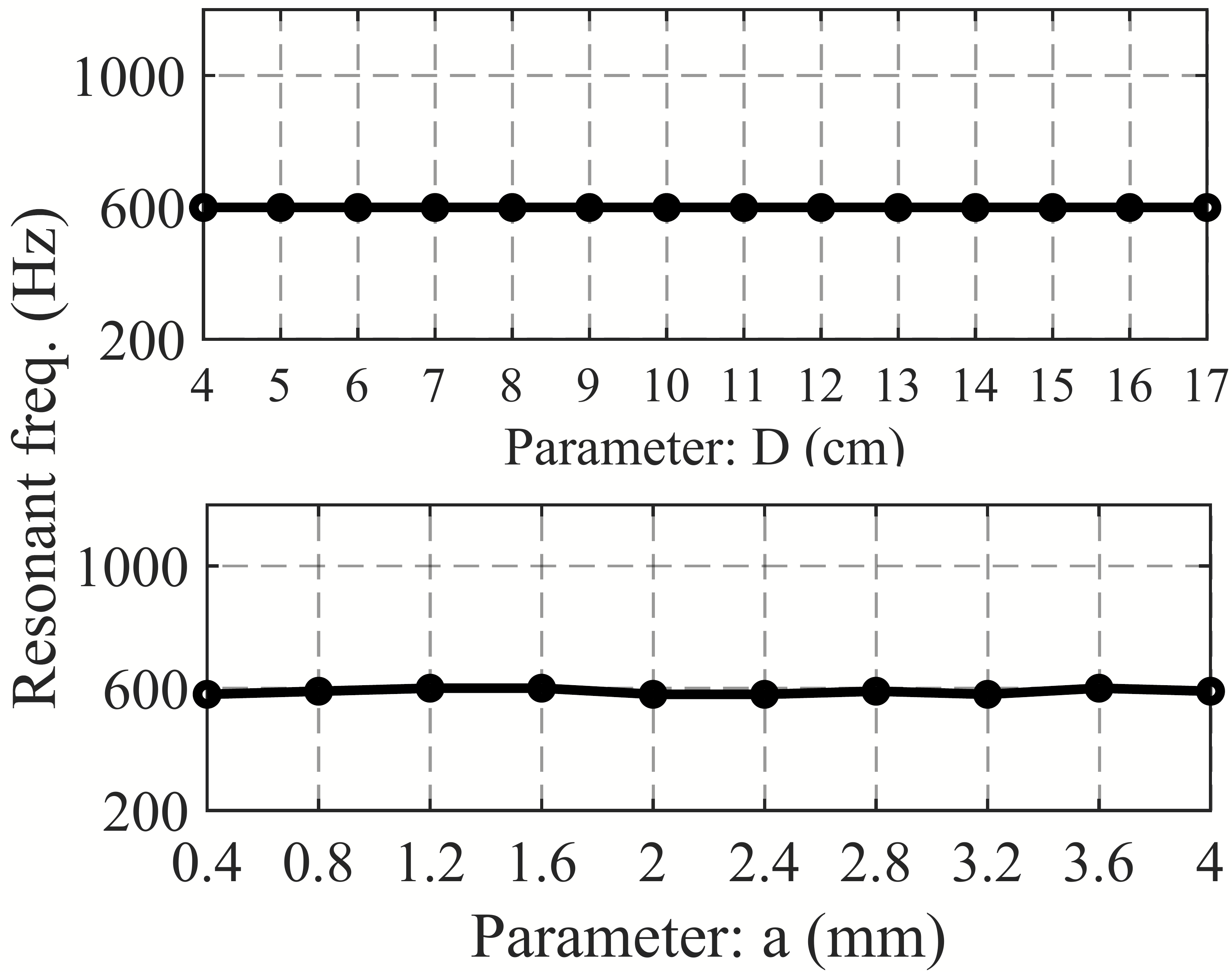}
        \label{aa}}
        \hfill
\subfloat[]{
		\includegraphics[scale=0.045]{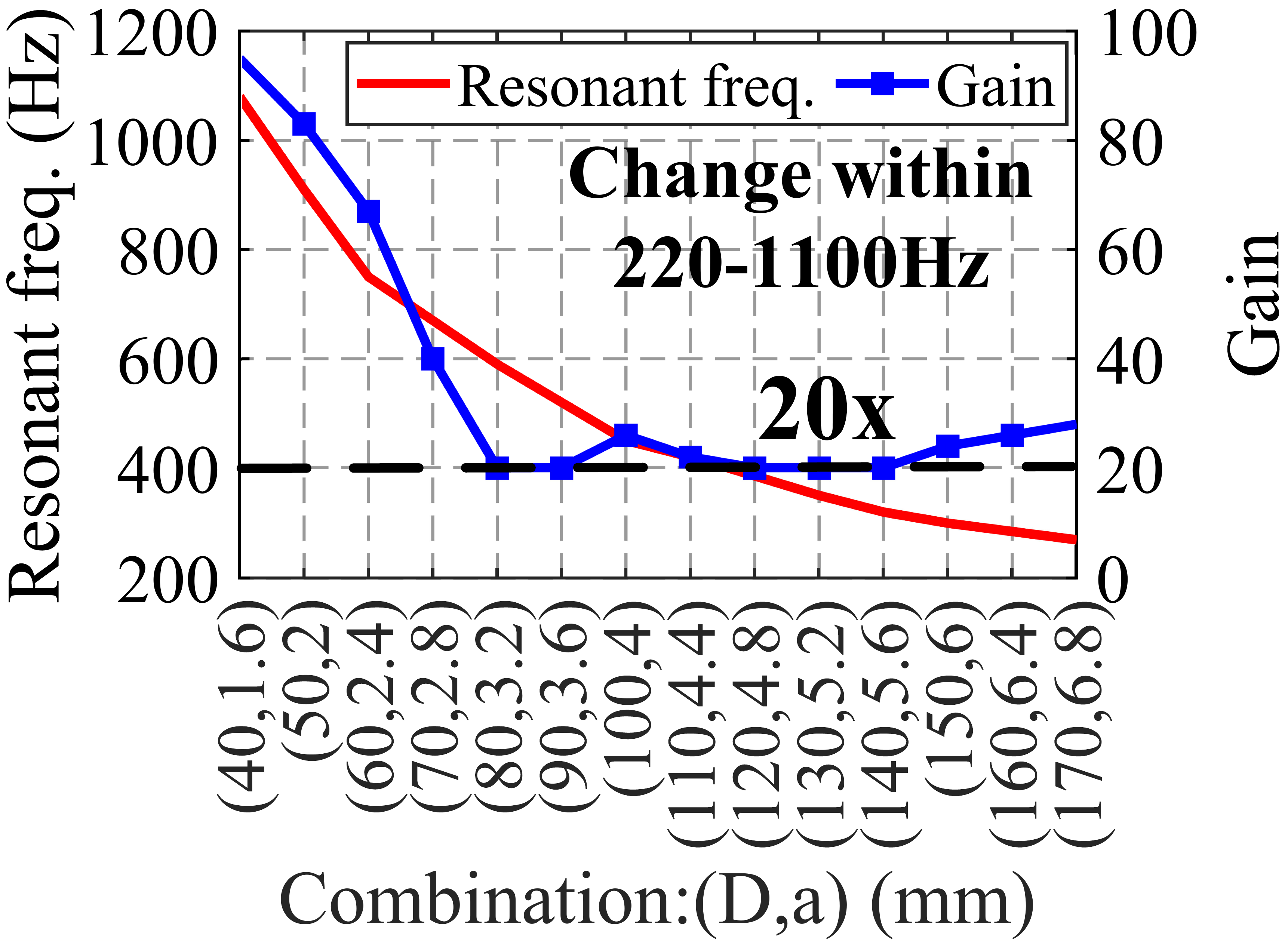}
        \label{Da}}
        \hfill
\subfloat[]{
		\includegraphics[scale=0.045]{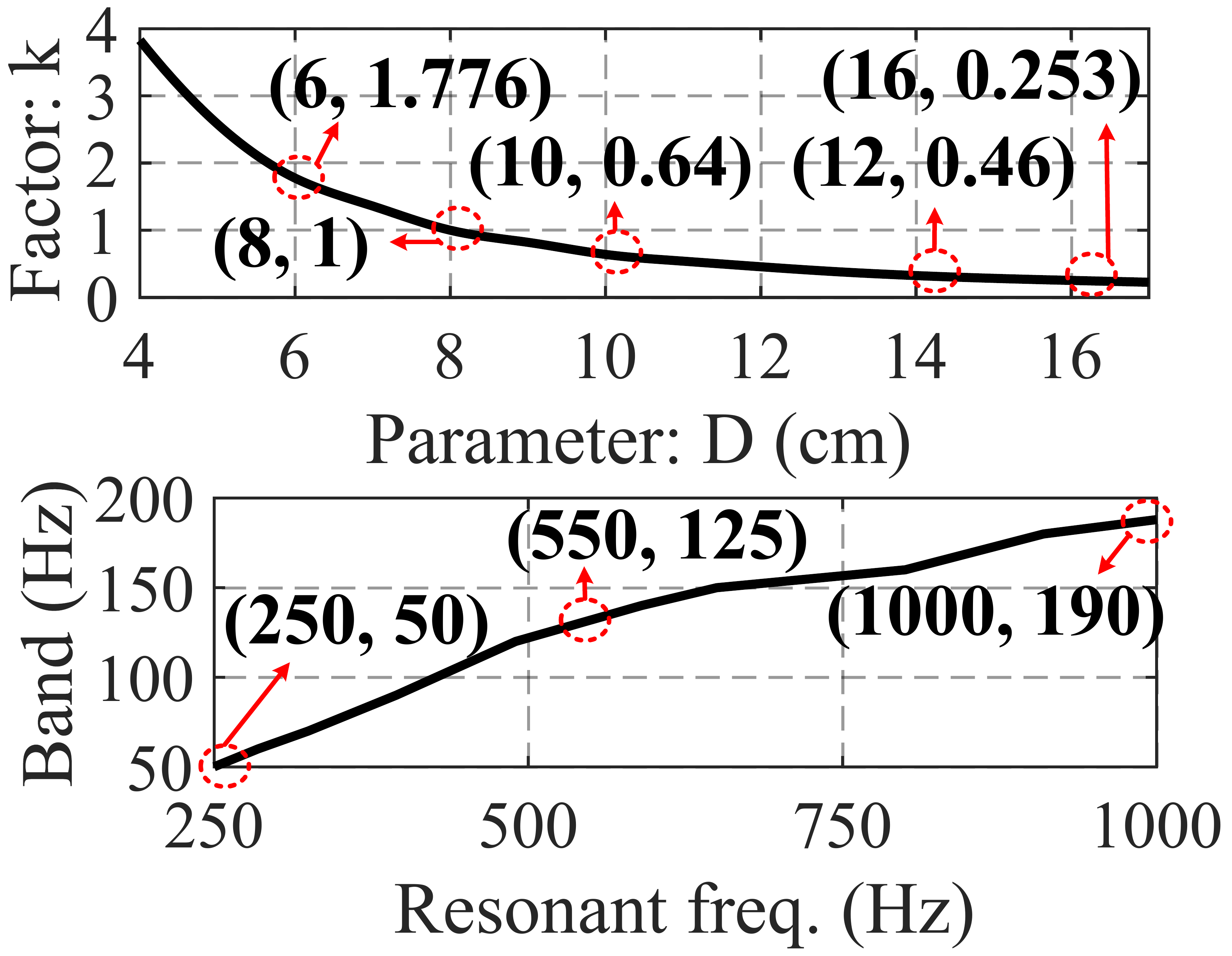}
        \label{KD}}
\caption{(a) \textit{D} and \textit{a} individually. (b) \textit{D} and \textit{a} simultaneously. (c) Top: $k$ vs. \textit{D}; Bottom: resonant frequency vs. bandwidth.}
\label{parameter change}
 \vspace{-10pt}
\end{figure}

Based on structural parameter analysis, we propose a multi-metamaterial system design that avoids increasing the number of channels (\(N\)) and instead tunes other key parameters to shift the resonant frequency.  

As shown in Fig.~\ref{structure}, the outer diameter (\(D\)) and channel width (\(a\)) directly influence the sound transmission path, thereby affecting the effective refractive index (\(n_r\)) and resulting sound pressure (\(P\)). In contrast, wall thickness (\(t\)) can be treated as fixed, since varying \(t\) is essentially equivalent to adjusting \(a\). The inner diameter (\(d\)) of the central cavity also has little effect, as amplification mainly occurs inside the coiled channels. Using COMSOL (a widely used multiphysics simulation platform known for its ability to accurately solve coupled acoustic-structural problems. It has been extensively validated in metamaterials and acoustics research, making its simulation results highly reliable~\cite{r11,r12,r14,r4}), we examined how adjusting \(D\) and \(a\) while keeping \(N\) fixed affects the resonant frequency (Fig.~\ref{parameter change}).  

When varied individually, changes in \(D\) show almost no influence, while changes in \(a\) shift the central frequency by only ~50 Hz (Fig.~\ref{aa}), neither of which meets the low-frequency enhancement requirement. However, simultaneous tuning of \(D\) and \(a\) produces a significant downward shift of the resonance (Fig.~\ref{Da}). The relationship can be expressed as:  
\[
f_c = k \times f_{c,0} \times \frac{(D , a)}{(D_0 , a_0)},
\]  
where \(f_{c,0}=563\ \text{Hz}, D_0=80\ \text{mm}, a_0=3.2\ \text{mm}\).  

The proportionality factor \(k\) decreases as \(D\) increases, showing nonlinear behavior: highly sensitive at small sizes and leveling off as size grows (Fig.~\ref{KD}). This relationship provides the foundation for designing multi-metamaterial systems. COMSOL simulations confirm that the new design achieves low-frequency gain exceeding 15×, while avoiding the thermo-viscous losses that limit traditional high-\(N\) resonators (Fig.~\ref{Da}).

\subsection{Expanding bandwidth via multi-metamaterials} \label{chap:6.2}

\begin{algorithm2e}[!t]
\footnotesize
\SetAlgoLined
\caption{\textit{Optimal Number of Metamaterials}}
\label{alg:metanum}
$Input:\ f_{start},\ f_{end},\ BW(f_c)$\\
\SetKwFunction{FB}{Optimal Design}
\SetKwFunction{Design}{DesignMetamaterials}
\SetKwFunction{Segment}{SegmentFrequency}
\FB{$f_{start}, f_{end}, BW(f_c)$}\\   
\SetKwProg{Fn}{Function}{:}{}
\Fn{\\ \FB{$f_{start}, f_{end}, BW(f_c)$}}{
    $resonant\_freqs \gets [\ ], f_{current} \gets f_{start}$ \;
    \While{$f_{current} < f_{end}$}{
            $f_c \gets f_{current}, bw \gets BW(f_c),  resonant\_freqs.\text{append}(f_c)$ \;
        $f_{current} \gets f_{current} + bw$ \;
    }
    $S \gets \text{len}(resonant\_freqs)$ \\
    \Return{$resonant\_freqs, S$}
}      

\end{algorithm2e}

\subsubsection{Defining the gain range}
To design an effective metamaterial system, we first identify the key frequency range for speech enhancement. While human speech spans from roughly 250 Hz to 4000 Hz, prior studies show that intelligibility depends most on frequencies below 1000 Hz~\cite{voicelistener,catford2001practical}. This is because vowel recognition, which is crucial for comprehension, relies on first formants (F1), which largely fall within 250  -  1000 Hz. We therefore target this range for enhancement.

\subsubsection{Determining the number of units.}
We then used COMSOL simulations to examine how metamaterial resonant frequencies relate to bandwidth (bottom of Fig.~\ref{KD}). Results show narrower bandwidths in the 250 - 550 Hz range (around 87.5 Hz) and wider bandwidths in the 550 - 1000 Hz range (around 157.5 Hz). This behavior can be explained by material properties: bandwidth scales with resonant frequency, $BW = \alpha \cdot f_c$, where $\alpha$ is a material constant. For our resin - based design, higher resonant frequencies yield broader bandwidths.  
To cover the 250 - 1000 Hz range, we apply Algorithm~\ref{alg:metanum}. Starting from $f_{start}=280$ Hz (bandwidth ~60 Hz), the algorithm iteratively selects resonant frequencies and their bandwidths until reaching $f_{end}=1000$ Hz. This yields eight distinct resonant frequencies, corresponding to eight metamaterial units ($S=8$).  
\SystemName therefore consists of eight acoustic metamaterial units (Fig.~\ref{structuresys}, Table~\ref{Parameter}). The thickness $h$ is fixed at 15 mm, sufficient to house a microphone, producing a total thickness of 120 mm. This compact design spans 250–1000 Hz effectively while maintaining portability.

\subsubsection{Spectrum cropping and stitching.}
Each metamaterial unit amplifies only a narrow band, so the outputs must be combined to achieve continuous coverage. Based on COMSOL simulations, we trimmed overlapping regions at frequencies with the highest gain (320, 360, 430, 520, 620, 700, and 860 Hz), retaining only ranges with $>5\times$ gain. Because each unit covers a different band, mutual interference is minimal. Stitching the trimmed outputs provides full coverage of the target speech range, enabling the system to reconstruct intelligible voice. The effectiveness of this design choice is validated in Sec.~\ref{B3}.

\begin{table}[t!]
    \scriptsize
    \caption{Parameter Combinations of Acoustic Metamaterials.}
    \label{Parameter}
    \vspace{1mm}
    \centering
    \setlength{\tabcolsep}{8pt} 
    \begin{tabular} {lllll}
    \toprule
    \textbf{label} & \textbf{path width (\textit{a})} & \textbf{Diameters (\textit{D})} & \textbf{Resonant freq.} & \textbf{Freq. range}\\
    \midrule
    (1) & 2 mm &  50 mm & 930 Hz&840 - 1000 Hz \\
    (2) & 2.4 mm &  60 mm & 790 Hz&720 - 860 Hz \\
    (3) & 2.8 mm &  70 mm & 670 Hz&580 - 700 Hz\\
    (4) & 3.2 mm &  80 mm & 563 Hz&490 - 597 Hz\\
    (5) & 4 mm   & 100 mm & 470 Hz&430 - 523 Hz\\
    (6) & 4.8 mm & 120 mm & 385 Hz &350 - 435 Hz \\
    (7) & 5.6 mm &  140 mm & 325 Hz &290 - 350 Hz\\
    (8) & 6.4 mm &  160 mm & 280 Hz &250 - 295 Hz\\  
    \bottomrule
    \end{tabular}

\end{table}

\begin{figure}[t]
    \centering
    \includegraphics[width=0.9\linewidth]{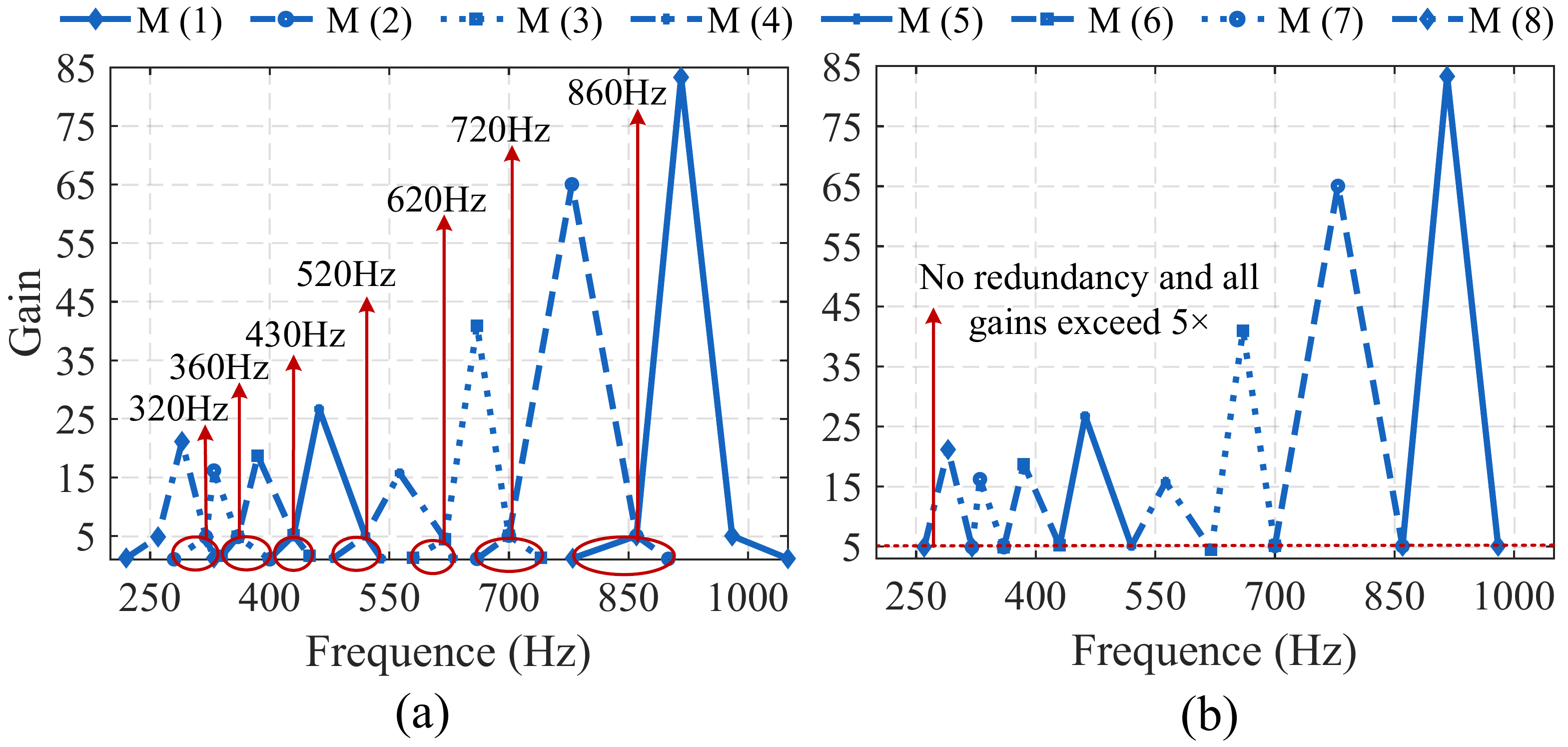}
\caption{(a) Gain curve of system and spectrum trimming (M: metamaterials), (b) Spectrum stitched after trimming.}
    \label{Enhancement curve}
    \vspace{-3mm}
\end{figure}

\subsection{Mitigating distortion and noise from metamaterial defects}
\label{chap:6.3}
Our optimizations described so far extend metamaterial across a wide speech range, but distortions and noise can occur. First, certain frequencies exhibit abrupt gain jumps ($>1500 \times$ as shown in Fig.~\ref{jumppoint}), typically caused by uneven material distribution from 3D printing defects. Second, the gain response varies across frequencies (Fig.~\ref{Enhancement curve}), which reduces audio fidelity and weakens eavesdropping performance. Third, metamaterials amplify both signals and noise within the same frequency bands, making outdoor deployment vulnerable to interference. We address these issues with algorithms for distortion suppression and noise reduction.

\subsubsection{Eliminating gain jumps}
To mitigate abrupt frequency jumps, we apply a smoothing algorithm that combines median filtering with threshold-based replacement. Gain jumps are detected by computing
\[
\Delta G(f_i) = \left| G(f_i) - G_{\text{median}}\left(f_{i-n}, f_i, f_{i+n}\right) \right|, 
\]
with \(n = 5,4,3,2,1\). Larger step sizes (5 Hz) capture wide-range anomalies, while smaller steps capture subtle local variations. Measurements across metamaterial samples show that all jumps occur within a 5 Hz window, consistent with geometric errors from 3D printing.  A threshold is then applied: if \(\Delta G(f_i)\) exceeds the mean plus three standard deviations, the value is replaced with the local median:  
\[
G_{\text{new}}(f_i) = G_{\text{median}}(f_{i-n}, f_i, f_{i+n}), \quad n=5,4,3,2,1.
\]
Iteratively applying this process removes outliers, producing a smooth gain curve (as illustrated in Fig.~\ref{jumppoint}) and a more coherent reconstructed signal.

\subsubsection{Maintain frequency gain balance}
\begin{figure}[!t]
  \centering
  \subfloat[Remove jumps in enhancement curve]{%
    \includegraphics[scale=0.14]{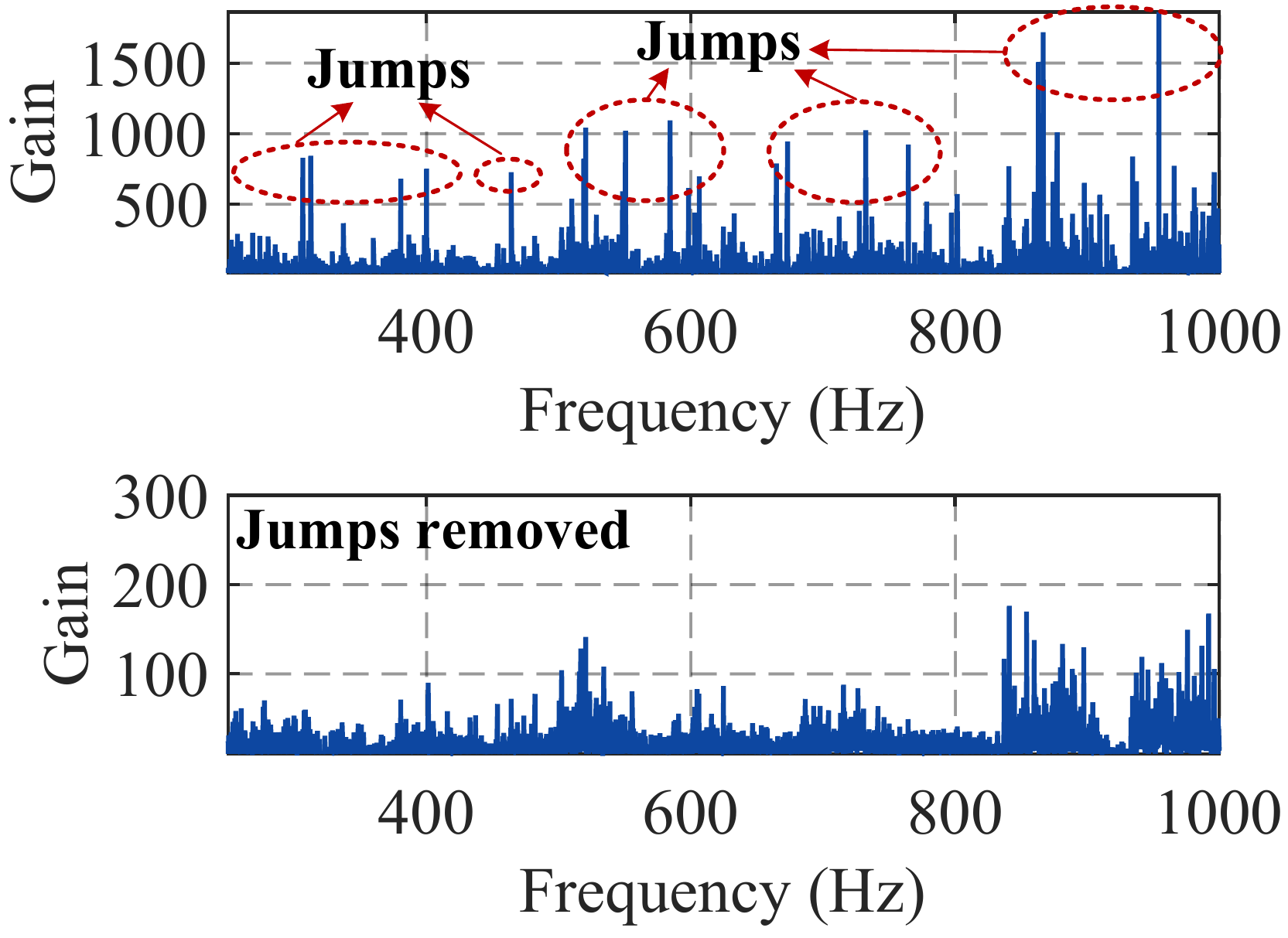}%
    \label{jumppoint}%
  }\hfill
  \subfloat[Enhancement curve after balancing]{%
    \includegraphics[scale=0.14]{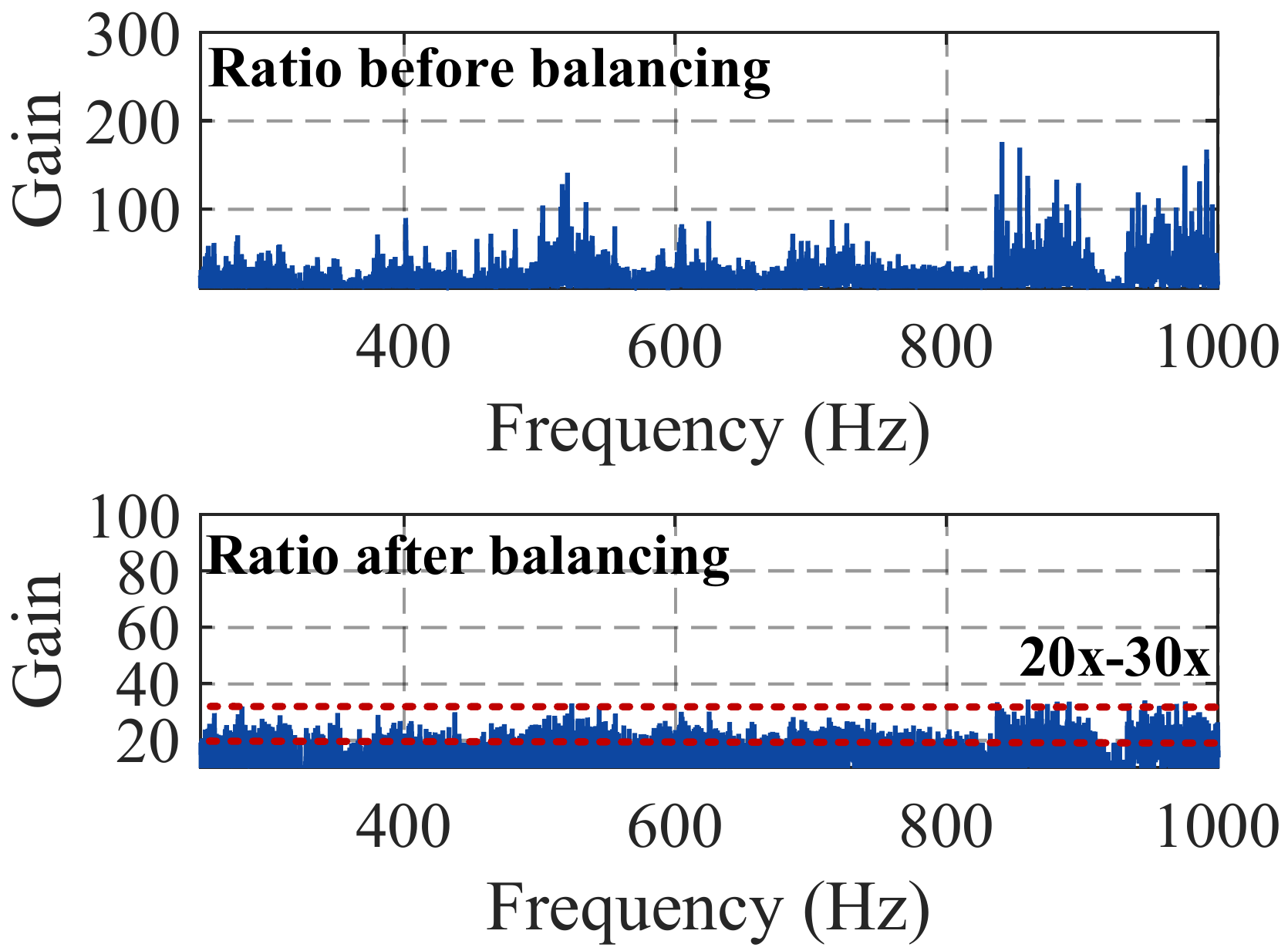}%
    \label{final}%
  }
  \caption{Optimization result: eliminated jumps in the enhancement curve and generated a new balanced curve.}
  \label{enhance}
      \vspace{-5mm}
\end{figure}

Even after removing jumps, gain mismatches across bands can degrade quality. To equalize performance, we implement an adaptive gain adjustment algorithm that dynamically scales each band relative to a target gain, set as the median of the curve. For excessively amplified bands:
$\text{new} = \frac{G_{\text{target}}}{G_{\text{actual}}}$,
and for under-amplified bands:
$\text{new} = \frac{G_{\text{actual}}}{G_{\text{target}}}$.
This balances all frequency responses, yielding a stable $20\times$ - $30\times$ gain across 250–1000 Hz (Fig.~\ref{final}). Together, smoothing and balancing form the \textit{Distortion Suppression Algorithm}, which improves both intelligibility and fidelity.


\subsubsection{Suppress noise interference}

\begin{algorithm2e}[!t]
\footnotesize
    \SetAlgoLined 
	\caption{\textit{Noise Suppression Algorithm}}
    \label{Algorithm2} 
    \SetKwFunction{FB}{Noise Suppression}
    \SetKwFunction{one}{Step1}
    \SetKwFunction{two}{Step2}
    \SetKwProg{Fn}{Function}{:}{}       
    $\FB{}$\\
    \label{alg2:b8}      
    \Fn{\FB{}}{
        $\one{};\two{$f_0$}; \textbf{return}\ \two{$f_0$} $\\
    }
    
    \Fn{\one{}}{
        $Receive\ noise\ and\ analyze\ frequency\ range\ f_0; \textbf{return}\ f_0$ \label{alg:b10} \\
    }

\Fn{\two{$f_0$}}{
    $S\ denotes\ the\ optimal\ number\ of\ metamaterials.$\\
    \For{$i = 1$ \textbf{to} S}{
        \If{$f_i$ contains $f_0$}{
            $f_i \gets \text{Isolate from the combined range}$ \\
        }}
        $Combine\ the\ remaining\ f_i\ into\ a\ new\ spectrum\ F; \textbf{return}\ F$
}              
\end{algorithm2e}

Since metamaterials amplify all sounds within their effective range, suppressing background noise is essential. We design a \textit{Noise Suppression Algorithm} (Algorithm~\ref{Algorithm2}) that detects noise-dominated frequency bands during spectrum synthesis and selectively suppresses the corresponding metamaterial outputs.  

Outdoor noise analysis shows that different types of noise exhibit distinct frequency distributions: environmental and traffic noise are typically concentrated in the 20–300 Hz range~\cite{300noiserange1,300noiserange2}, while industrial machinery noise mainly falls within 50–500 Hz~\cite{500noiserange1,500noiserange2}. These ranges only partially overlap with \SystemName’s 250 - 1000 Hz enhancement band, making selective suppression feasible. Our noise suppression algorithm first measures the environment to identify the dominant noise frequency range (\(f_0\)), then checks each metamaterial unit’s band (\(f_i\)) and excludes those overlapping with \(f_0\). The remaining bands are combined into a new spectrum \(F\), which reduces noise while preserving intelligible speech.  

When interference occurs within the enhanced range, \SystemName automatically reduces the gain of the affected channels to minimize noise contribution. Although this sacrifices some frequency coverage, experiments (Sec.~\ref{C2}) show that speech intelligibility is largely preserved while background interference is significantly reduced.



\subsection{Implementation} \label{chap:5}

\begin{figure}[!t]
  \centering
  \subfloat[Prototype of \SystemName and its flow diagram]{%
    \includegraphics[scale=0.24]{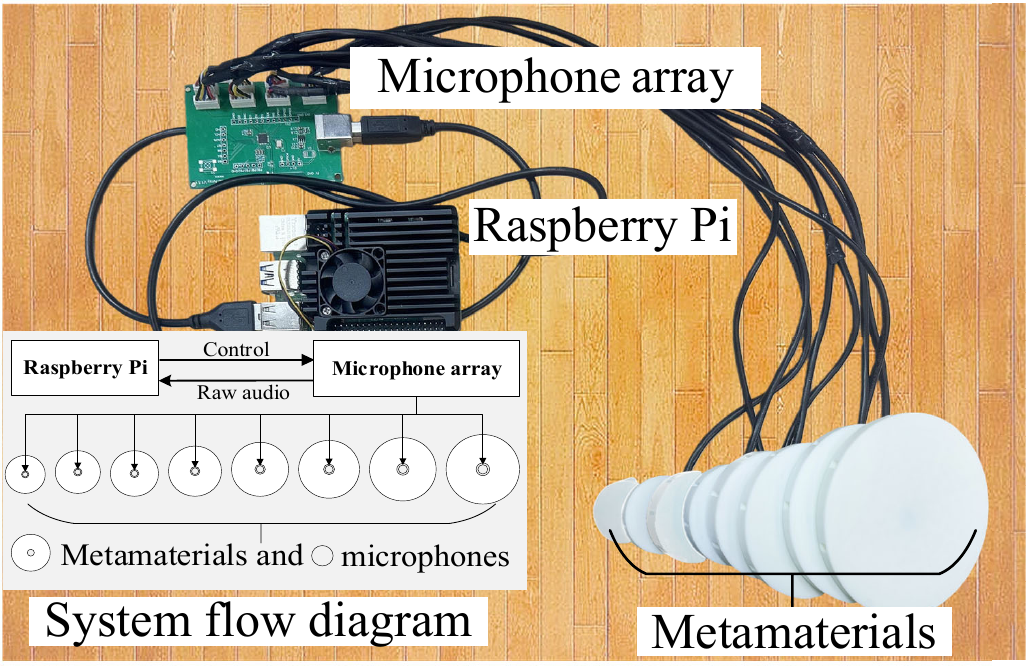}
    \label{prototype2}
  }\hfill
  \subfloat[Covert prototype (smaller than a 13-inch MacBook Air)]{%
    \includegraphics[scale=0.26]{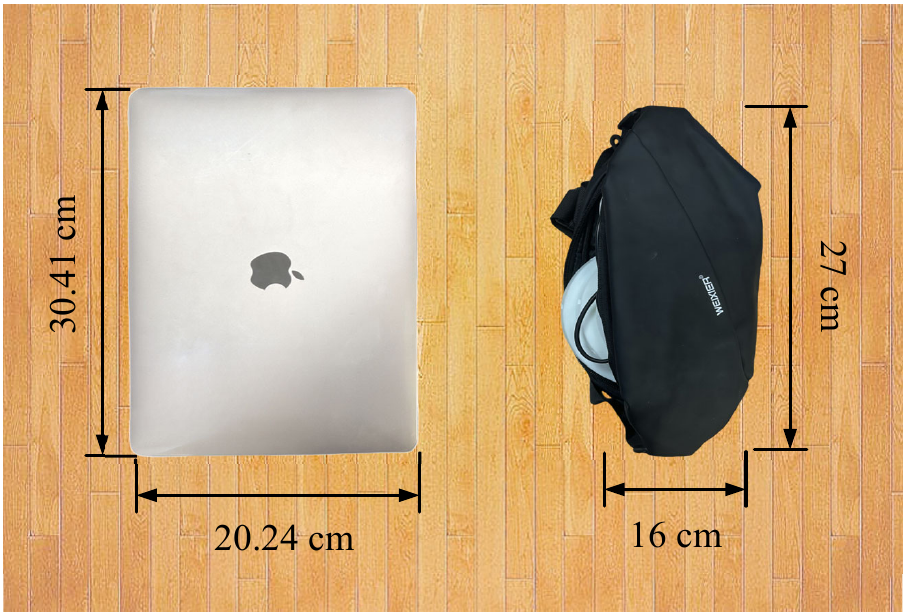}
    \label{covert2}
  }
  \caption{Implementation: prototype and covert form factor.}
  \label{fig:implementation}
      \vspace{-5mm}
\end{figure}
As shown in Fig.~\ref{prototype2}, the \SystemName prototype integrates eight acoustic metamaterial modules, a commercial microphone array, and a Raspberry Pi 5B, which performs on-device beamforming, MVDR filtering, noise suppression, and audio reconstruction. The assembled device fits in a ($27\times16$) cm handbag (Fig.~\ref{covert2}), enabling portable and covert operation in mobile attack scenarios.

\section{EVALUATION} \label{chap:6}
\subsection{Experimental Setup}
All our experiments were conducted under approval from the Institutional Review Board (IRB). The research equipment was self-funded, and participants voluntarily joined with informed consent. No sensitive or personally identifying data was collected or stored during the study, ensuring compliance with ethical standards.  

\begin{figure}[!t]
  \centering
  \subfloat[S1: environmental noise and obstacles]{%
    \includegraphics[scale=0.415]{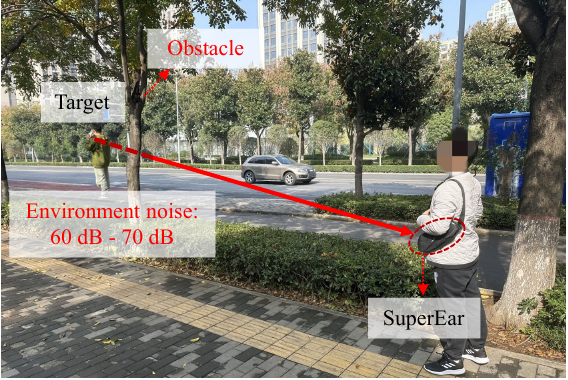}%
    \label{S2}%
  }\hfill
  \subfloat[S2: walking targets with wind]{%
    \includegraphics[scale=0.415]{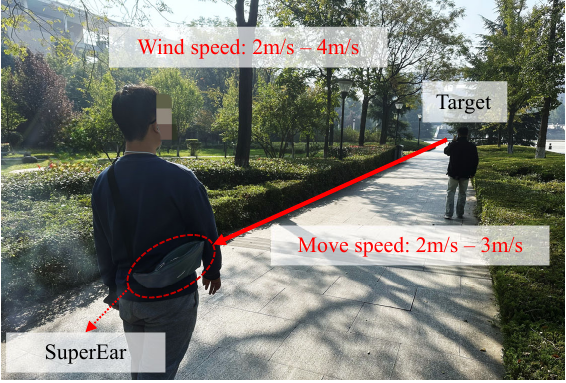}%
    \label{S1}%
  }
  \caption{Real attack evaluation scenarios: S1 and S2.}
  \label{scenarios}
\end{figure}

\begin{table}[t!]
    \scriptsize
    \caption{Mobile devices used in evaluation.}
    \label{Victim devices}
    \vspace{1mm}
    \centering
    \setlength{\tabcolsep}{21pt} 
    \begin{tabular} {p{1.3cm}lllll}
    \toprule
    \makecell[l]{\textbf{Brand}} &  \textbf{Model} & \textbf{OS}\\
    \midrule
    \rowcolor{gray!20}  & iPhone 16 Pro Max & IOS 18.4.1 \\
    \rowcolor{gray!20} \makecell[l]{\multirowcell{-2}{Apple~\cite{Apple}}}  & iPhone 15 Pro &  IOS 17.5\\
    \makecell[l]{\multirowcell{1}{HONOR~\cite{HONOR}}} & Magic V Flip & MagicOS 8 \\
   \rowcolor{gray!20} \makecell[l]{\multirowcell{1}{Samsung~\cite{Samsung}}} &  Galaxy Z Fold 4  & Android 14 \\  
      \makecell[l]{Google~\cite{Google}}& Pixel 8 Pro &  Android 14\\
     \rowcolor{gray!20} \makecell[l]{Sony~\cite{Sony}}&  Xperia 10 IV  &  Android 13 \\  
    & Xiaomi 14  &  HyperOS 2 \\    
   \makecell[l]{\multirowcell{-2}{Xiaomi~\cite{Xiaomi}}} & Redmi K50 Ultra & Miui 13.0.1 \\     
\rowcolor{gray!20} \makecell[l]{\multirowcell{1}{Huawei~\cite{Huawei}}} & Mate 60 Pro & HarmonyOS 4\\
    \bottomrule
    \end{tabular}
 \vspace{-5mm}
\end{table}

\subsubsection{Test targets}
As given in Table~\ref{Victim devices}, we evaluated \SystemName against nine mainstream smartphones. In addition, three volunteers (one female, two males) acted as callers to test the system’s ability to capture spoken content. We also recruited 10 volunteers (5 female and 5 males) to evaluate the similarity between the original voice and the one captured by \SystemName. All our volunteers are postgraduate students studying at our university. 

\subsubsection{Metrics}
Following prior work~\cite{Mel,mmecho,AccelEve,Vibphone,mmspy,accear}, we use four metrics: \emph{success rate}, \emph{range of successful attack (RSA)}, \emph{word accuracy}, and \emph{Mean Opinion Score (MOS)}. Success rate is computed over 30 trials per device-condition pair, with reconstructed audio considered successful if Mel-Cepstral Distortion (MCD, lower is better) < 8~\cite{Mel,mmecho}. RSA is the maximum distance with success rate $\geq$ 80\%~\cite{AccelEve}. Word accuracy is the fraction of correctly recognized words via Google Speech Recognition~\cite{Google,AccelEve,Vibphone,mmspy}. MOS is rated by 10 gender-balanced volunteers on a 1–5 scale~\cite{mmecho,accear}.

\subsubsection{Experiment design}
Our experiments simulate realistic phone-call scenarios. In each trial, the target device played a scripted phrase at about 65 dB, while the volunteer repeated it at a whisper level of roughly 45 dB.

\cparagraph{Roadmap}
Our evaluation is organized into four parts: baseline performance (Sec.~\ref{sec:ep}), design choices and ablations (Sec.~\ref{sec:ecs}), real-world scenarios (Sec.~\ref{sec:prw}), and comparison with prior methods (Sec.~\ref{D}). All experiments were repeated multiple times, and mean results are reported.

\subsection{Eavesdropping Performance of \SystemName\label{sec:ep}}
\begin{figure} [t!]

    \centering
    \includegraphics[width=0.95\linewidth]{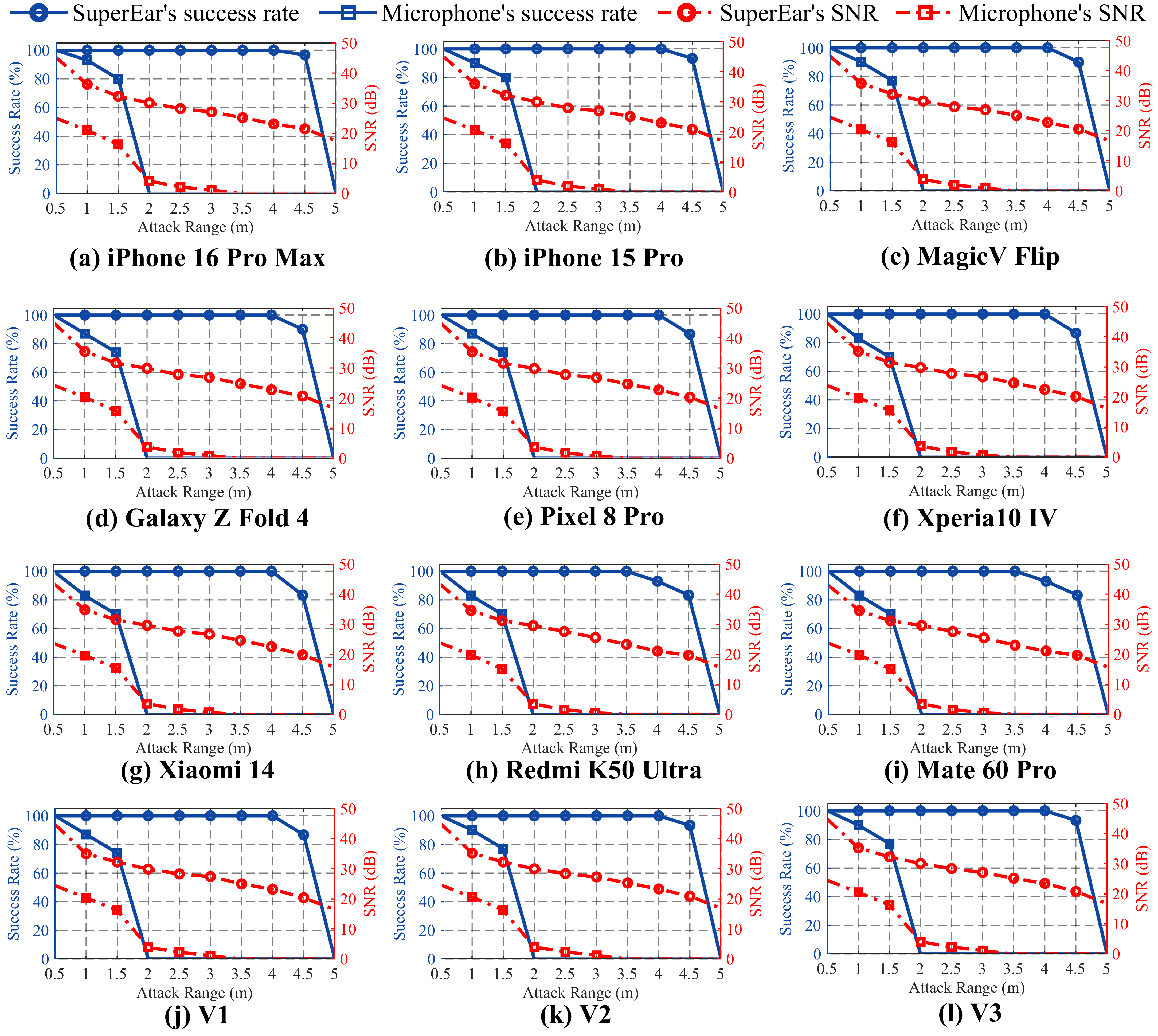}
    \caption{Performance of \SystemName and microphone array.}
    \label{devices and range}
    \vspace{-10pt}
\end{figure}

\begin{figure} [t!]
    \centering
    \includegraphics[width=0.95\linewidth]{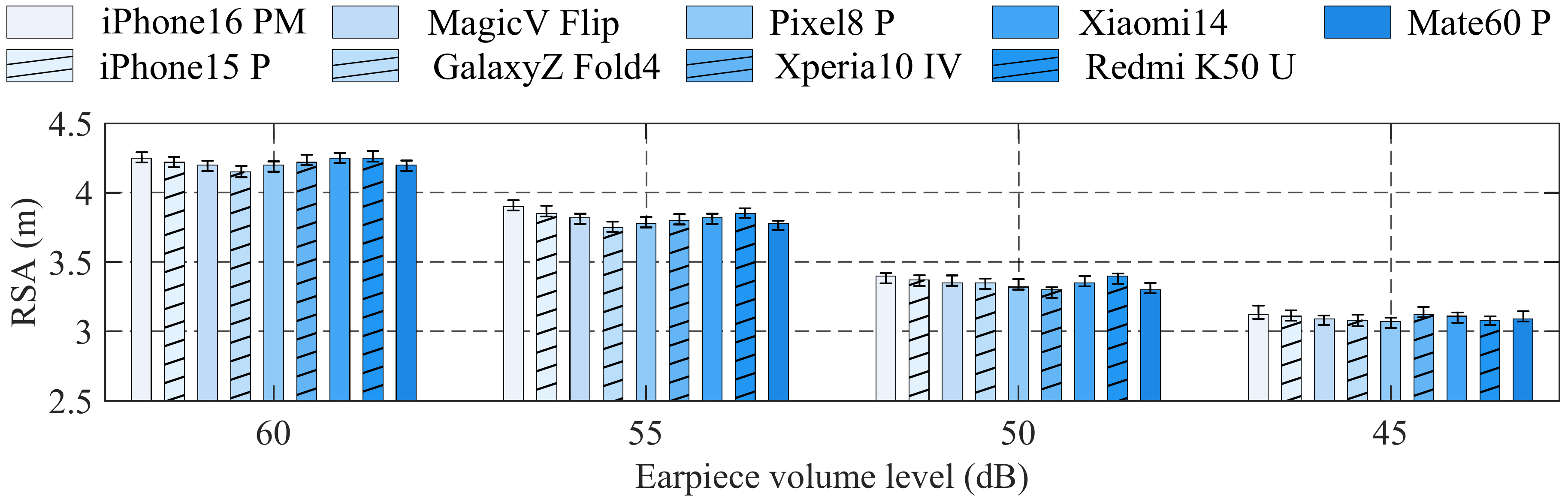}
    \caption{Performance under different volumes (covering four typical and low earpiece volume levels across different devices).}
    \label{volume}
     \vspace{-10pt}
\end{figure}

\begin{figure} [t!]
    \centering
    \includegraphics[width=0.95\linewidth]{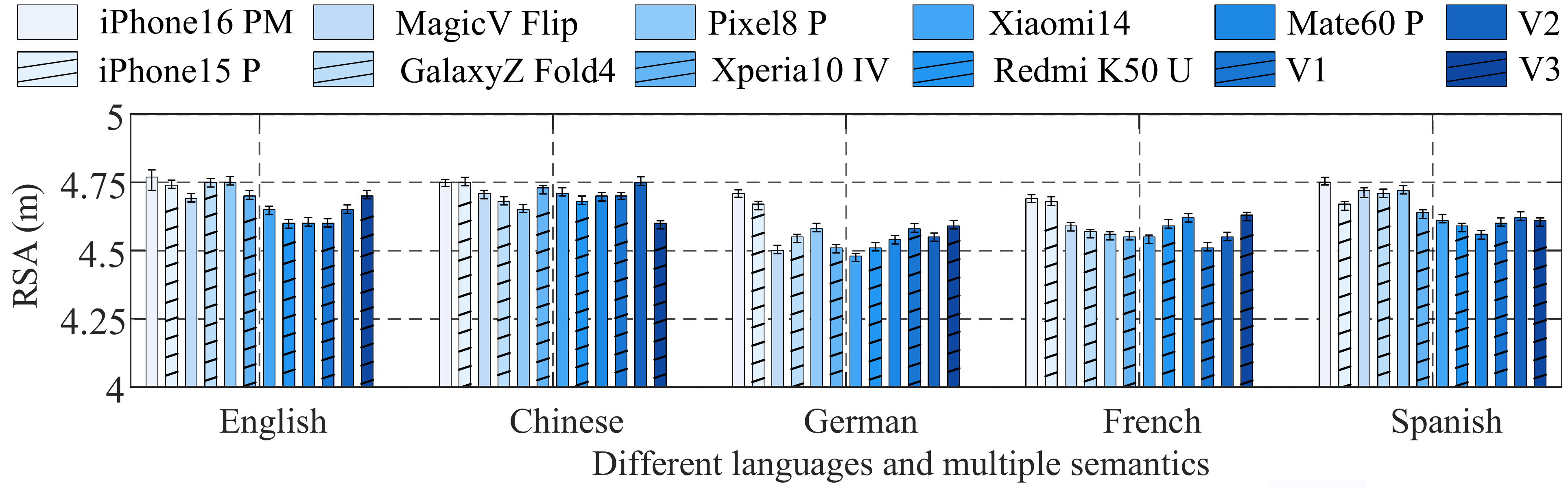}
    \caption{Performance under different languages and semantics (covering five languages and various semantic scenarios).}
    \label{language}
     \vspace{-10pt}
\end{figure}

\subsubsection{Eavesdropping range for different targets}\label{A1}
Eavesdropping range is a key metric for evaluating attack effectiveness. We measured the SNR of reconstructed audio across multiple devices and volunteers (Table~\ref{Victim devices}). Results (Fig.~\ref{devices and range}) show that \SystemName consistently achieves 4.6 meters on different devices and individuals, with success rates over 80\% and average SNRs of 20.4 dB and 20.7 dB, demonstrating strong stability and adaptability. In contrast, the average maximum eavesdropping range of microphone arrays is only 1.3 meters~\cite{mmeve,earspy}, far below the \textit{safe distance}.

\subsubsection{Effect of earpiece volume}\label{A2}
Earpiece volume directly influences sound energy and propagation distance, thereby affecting eavesdropping performance. Building on the 65 dB setting, we evaluated lower volumes of 60, 55, 50, and 45 dB. As shown in Fig.~\ref{volume}, even at approximately 45 dB, \SystemName maintains a stable eavesdropping range of about 3.2 m, exceeding the safe distance. This demonstrates that effective eavesdropping is possible even during low-volume calls. Since typical call volumes are generally higher, \SystemName remains practical in real-world scenarios.

\subsubsection{Performance across different languages and multi-semantic content}\label{A3}
To evaluate multilingual performance, we tested English, Chinese, German, French, and Spanish with five privacy-sensitive phrases. As shown in Fig.~\ref{language}, the average eavesdropping distance for all languages exceeds 4.55 m, with English and Chinese performing best. Overall, \SystemName maintains stable performance across different languages and semantic content, demonstrating strong cross-language robustness.

\subsubsection{Human assessment of eavesdropping performance}\label{A4}
Our ten volunteers rated the reconstructed audio for multiple targets (Fig.~\ref{Subjective assessment}), with average MOS scores above 4, indicating most of the original speech was successfully recovered and verifying the excellent eavesdropping capability of \SystemName.

\subsubsection{Eavesdropping accuracy on passwords}\label{A5}
\begin{figure}[!t]
\centering
\subfloat[]{
		\includegraphics[scale=0.06]{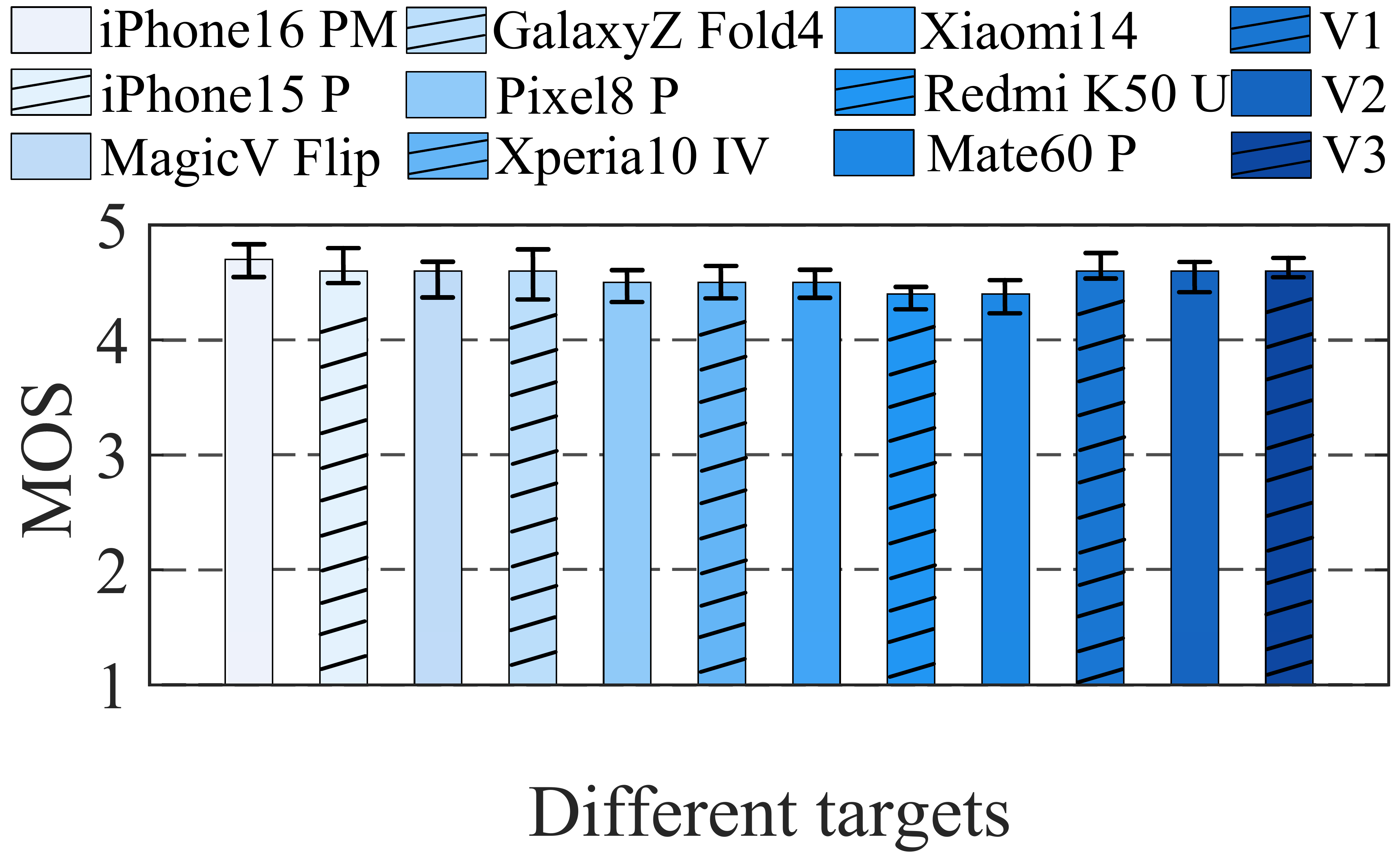}
  \label{Subjective assessment}}
\hspace{0.1cm}
\subfloat[]{
		\includegraphics[scale=0.118]{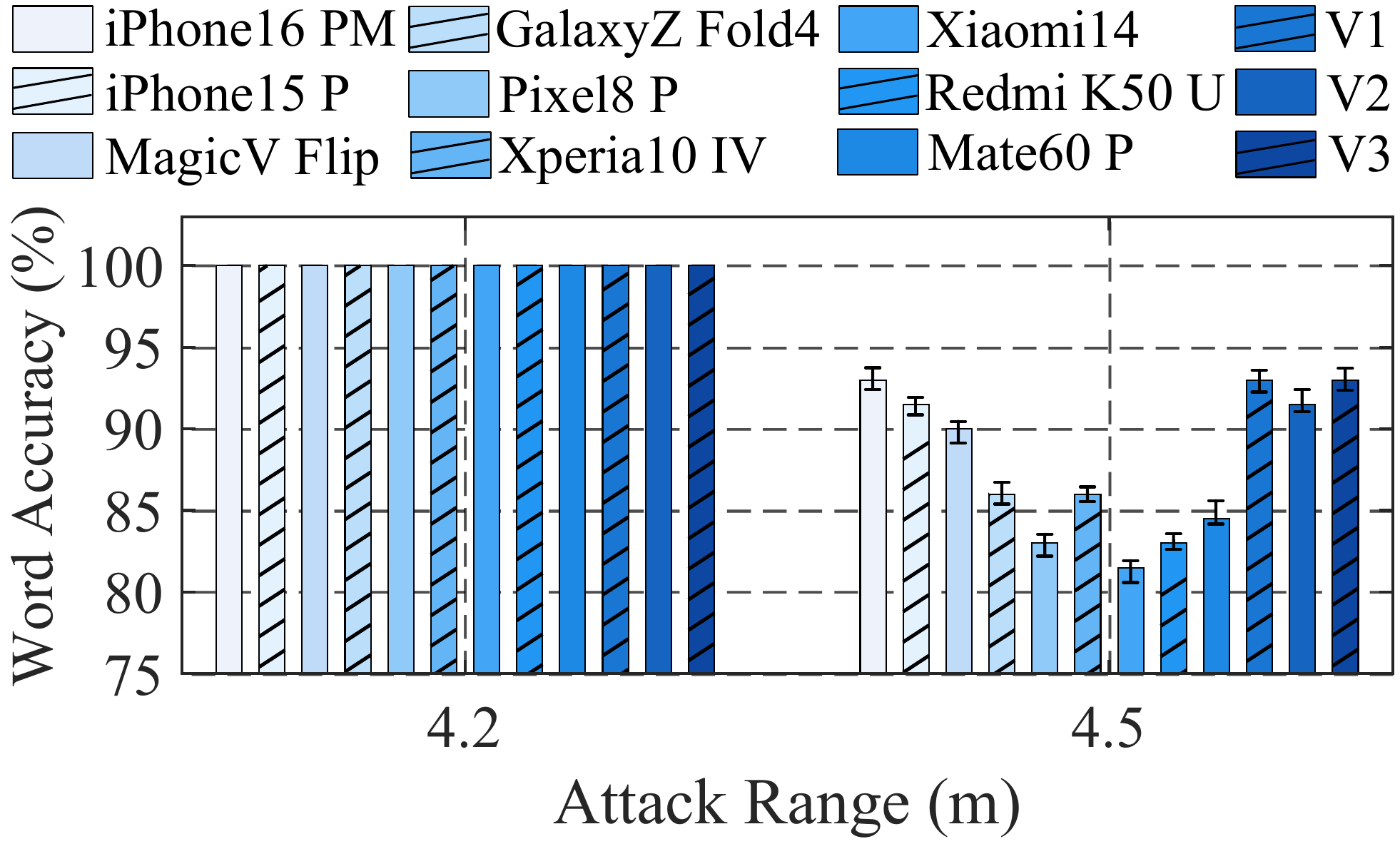}
  \label{acc}}
\caption{(a) Human auditory evaluation performance. (b) Eavesdropping accuracy against passwords.}
\label{1314}
\vspace{-10pt}
\end{figure}

Password eavesdropping is a risky method of information theft. The success of such an attack depends on whether \SystemName can accurately reconstruct each password digit. Evaluations show (Fig.~\ref{acc}) that at a distance of 4.2 meters, \SystemName achieves 100\% password recognition accuracy, demonstrating its high precision in remote eavesdropping.

\subsubsection{Performance with electret microphones}\label{A6}
\begin{figure}[!t]
\centering
\subfloat[]{
		\includegraphics[scale=0.06]{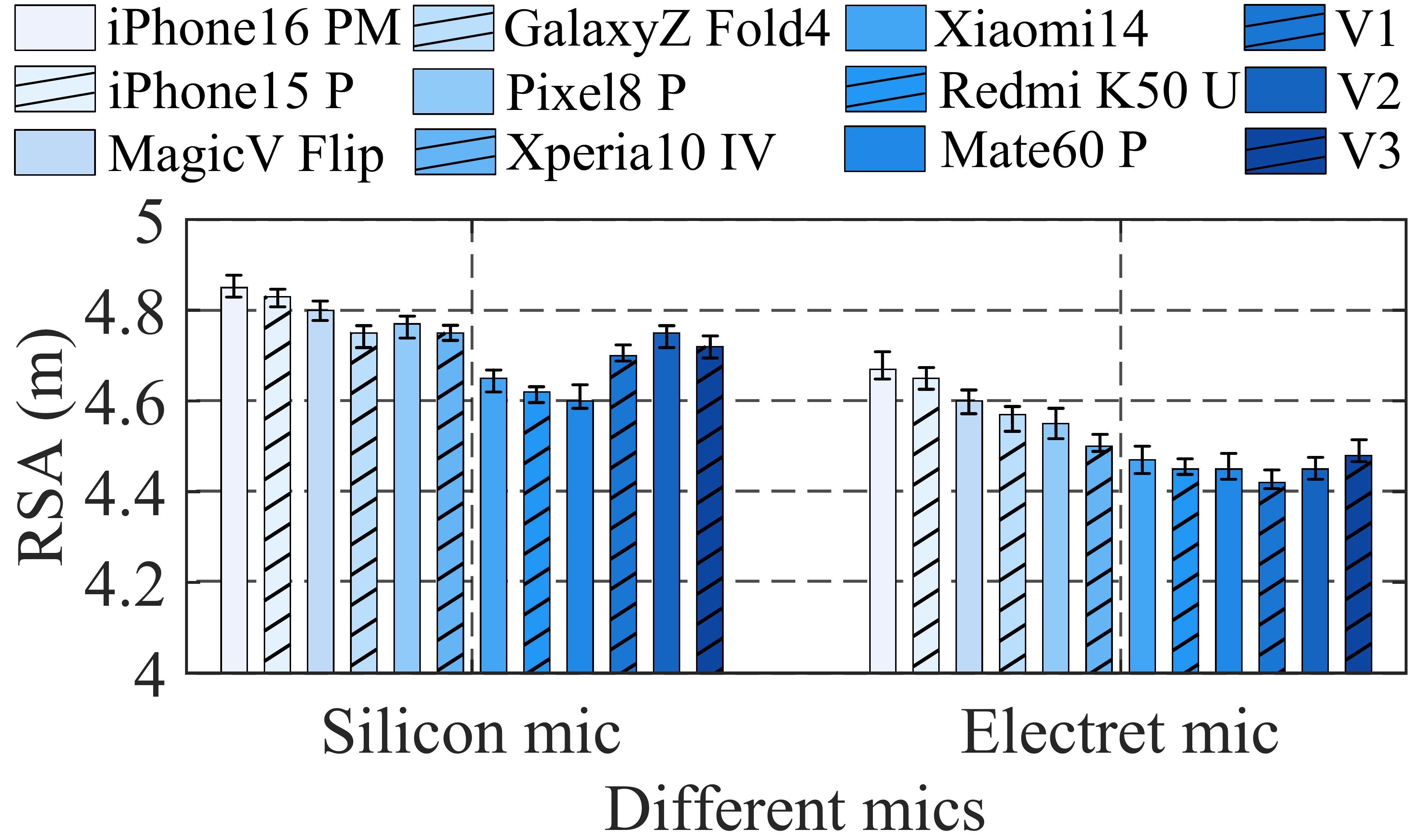}
  \label{differentmicrophones}}
\hspace{0.1cm}
\subfloat[]{
		\includegraphics[scale=0.055]{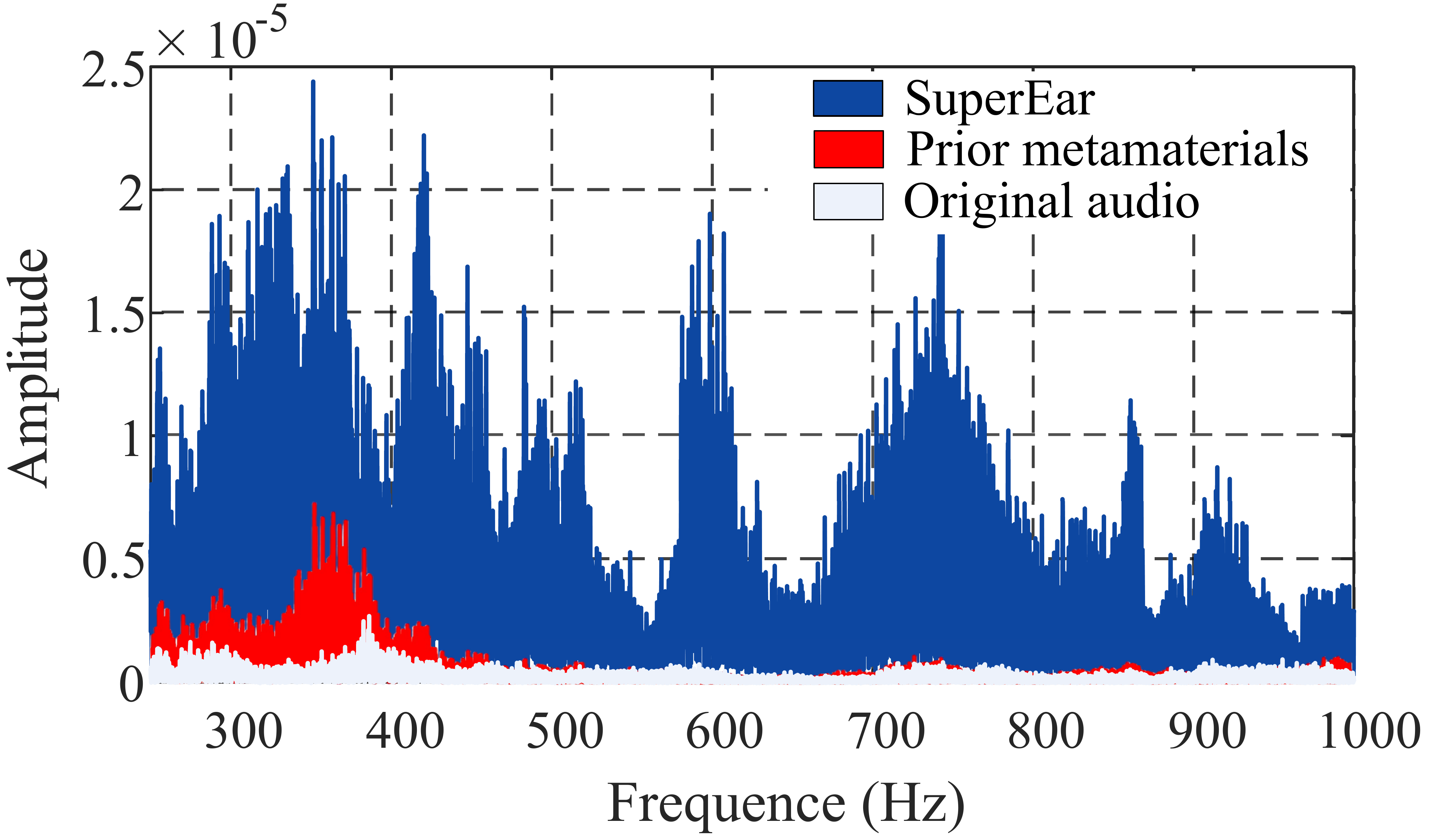}
  \label{real performance}}
\caption{(a) Performance with different microphones. (b) The real enhancement effect of \SystemName.}
\label{1516}
\vspace{-10pt}
\end{figure}

Electret microphones have slightly lower sensitivity but are low-cost and have a wide frequency response. After integration into \SystemName (Fig.~\ref{differentmicrophones}), the average RSA reached 4.52 meters. The metamaterial enhancement effectively compensates for the reduced sensitivity, maintaining eavesdropping performance.



\subsection{Enhanced Capability of \SystemName\label{sec:ecs}}

\subsubsection{Actual gain performance} \label{B1}
To evaluate \SystemName’s actual gain, we conducted laboratory experiments at a distance of 4.6 m under 43 dB background noise and without obstacles. As shown in Fig.~\ref{real performance}, \SystemName achieves a 20–30× gain in the 250–1000 Hz frequency range, closely matching simulation results, whereas prior low-frequency metamaterials provide only about 5× gain.

\subsubsection{Multi-metamaterial system performance} \label{B2}
The proposed multi-metamaterial system plays a critical role in long-range eavesdropping. Ablation experiments show that \SystemName achieves an average RSA of 4.6 m, representing an 86\% improvement over prior metamaterials with approximately 5× gain~\cite{ultra}. This performance far exceeds that of microphones and previous metamaterial designs, highlighting the system’s gain advantage.

\subsubsection{Distortion suppression algorithm performance} \label{B3}
The \textit{Distortion Suppression Algorithm} mitigates distortions caused by manufacturing imperfections and gain imbalance across multiple metamaterials. Without this algorithm, the system achieves an RSA of only 2.94 m, whereas \SystemName improves RSA by 53\%, demonstrating that the algorithm is essential for maintaining a safe eavesdropping distance.


\begin{figure}[!t]
\centering
\subfloat[]{
		\includegraphics[scale=0.115]{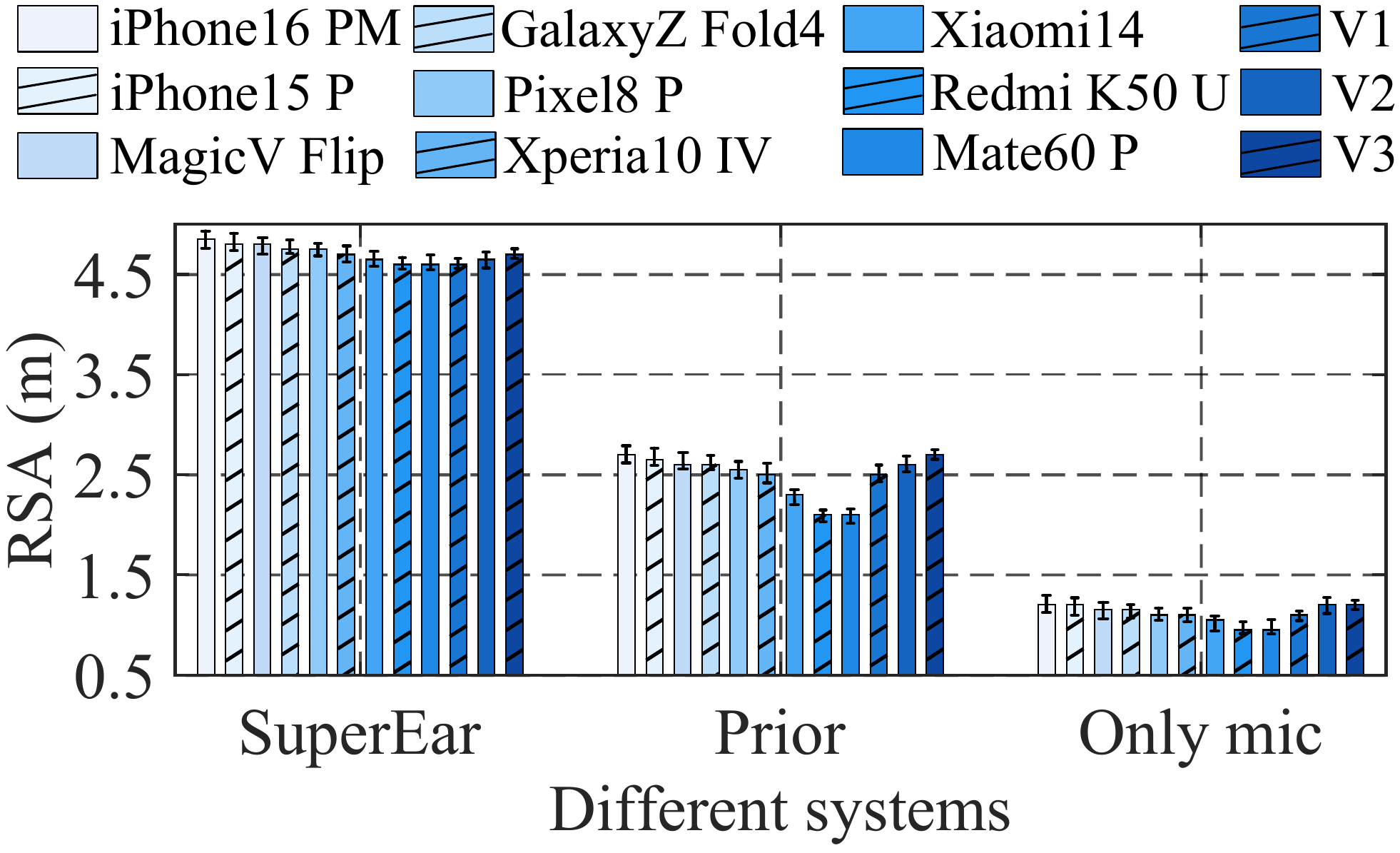}
  \label{enhancement capability}}
\hspace{0.1cm}
\subfloat[]{
		\includegraphics[scale=0.115]{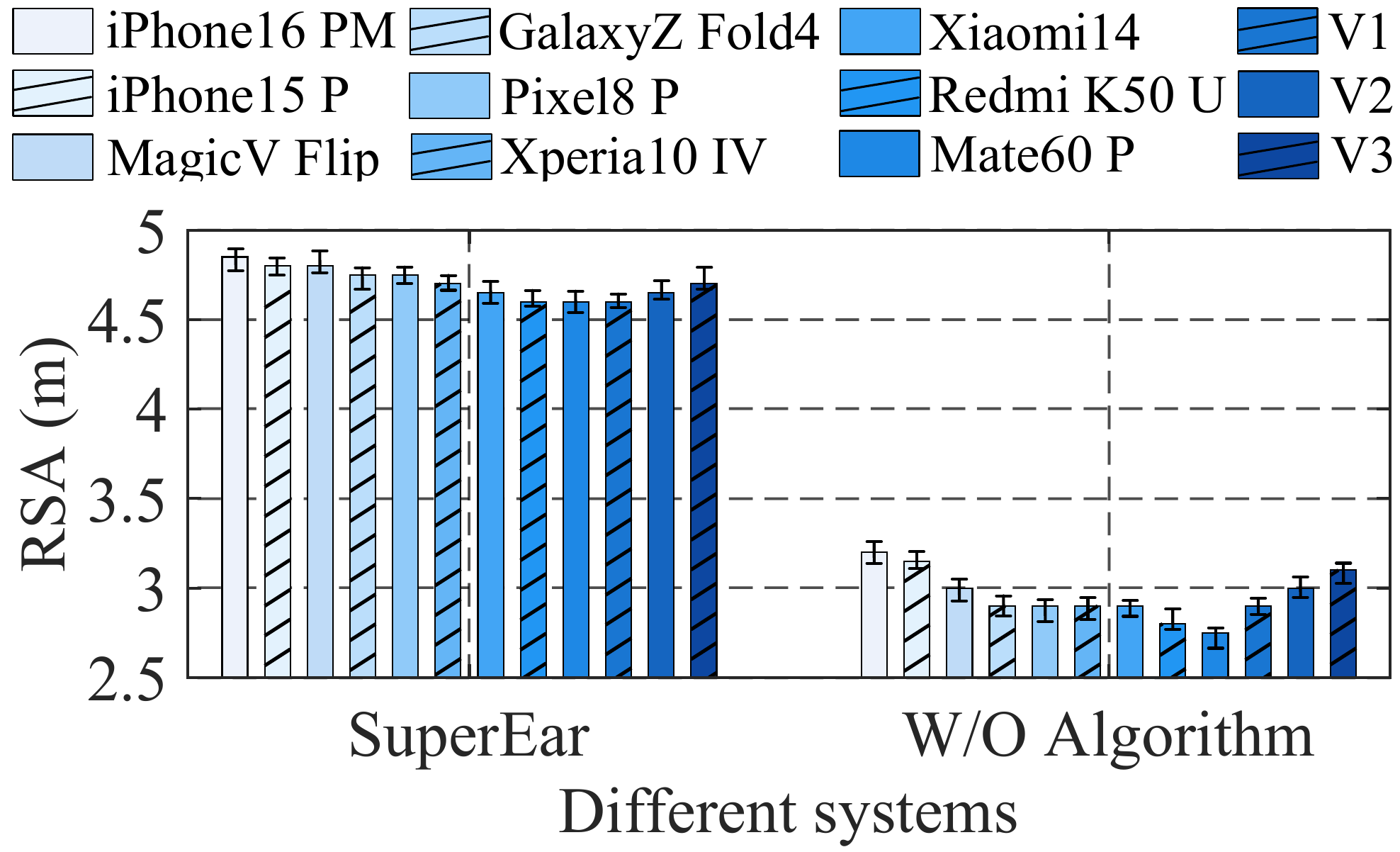}
  \label{WOA}}
\caption{(a) Improved metamaterial capability. (b) Performance of \textit{Distortion Suppression Algorithm.}}
\label{1718}
\vspace{-10pt}
\end{figure}

\begin{figure}[t!]
    \centering   
        \centering        \includegraphics[width=0.95\linewidth]{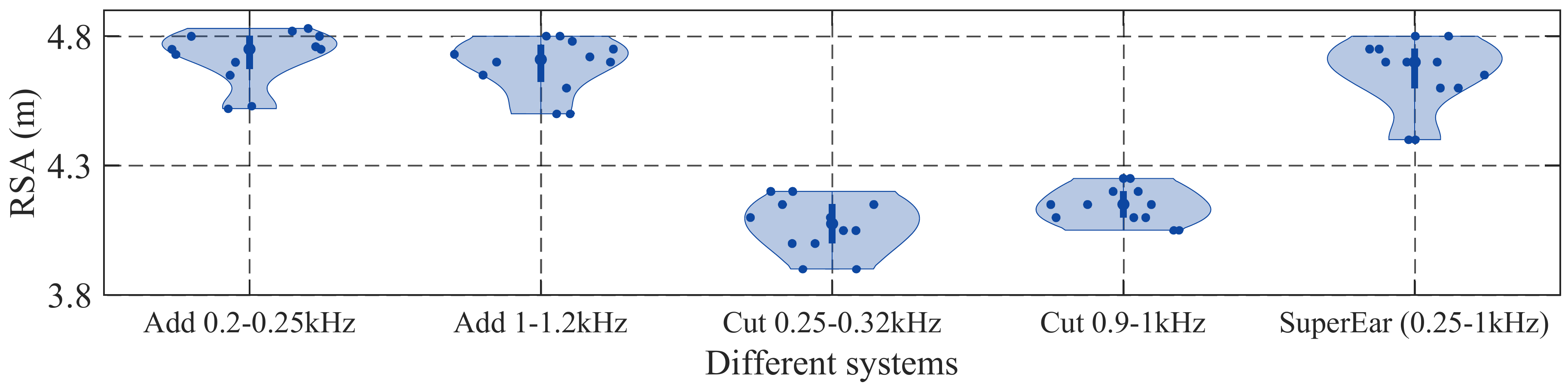}
        \caption{\SystemName's optimal metamaterials (covering the impact of adding or removing metamaterial units on the system).}
        \label{diffdesign}
      \vspace{-10pt}
 \end{figure}

\begin{figure}[t!]
    \centering   
        \centering        \includegraphics[width=0.95\linewidth]{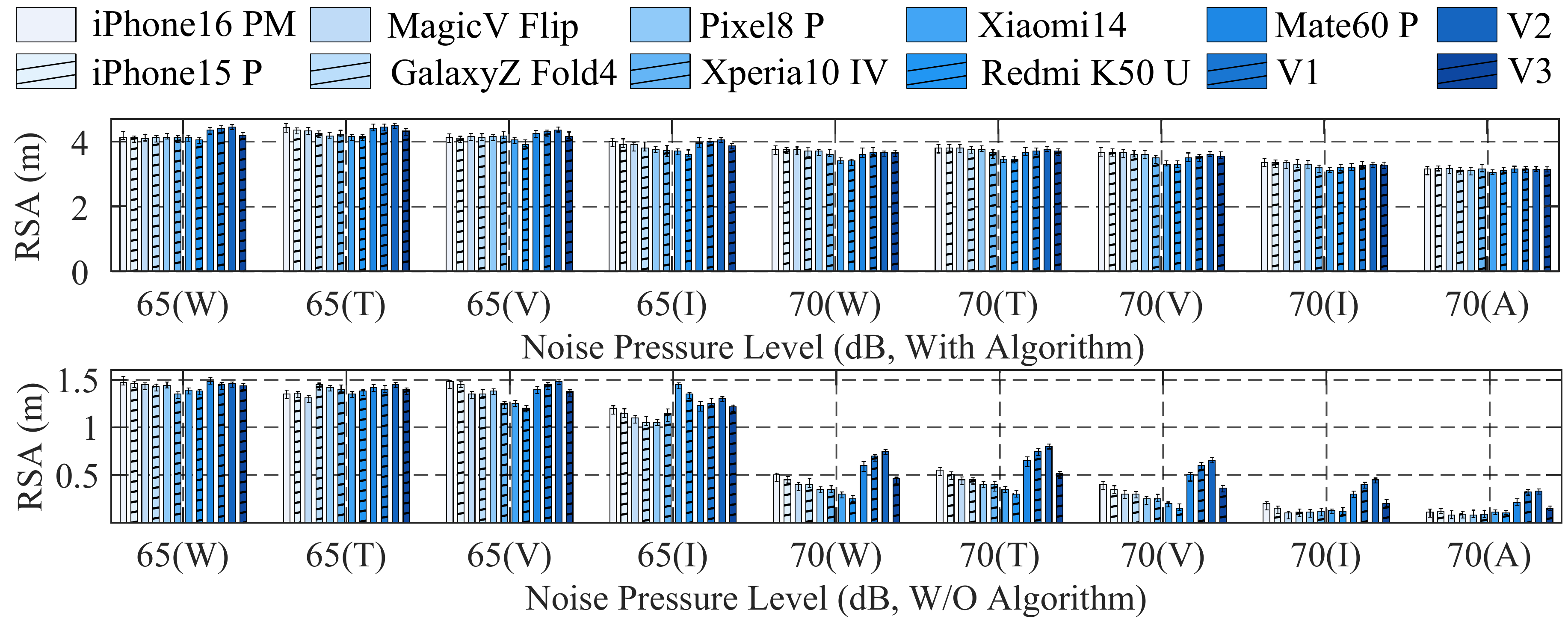}
        \caption{Noise impact (W: wind; T: thunder; V: vehicle; I: industrial; A: combination of the above noises).}
        \label{noise}
     \vspace{-10pt}
 \end{figure}

\begin{figure}[t!]
    \centering   
        \centering        \includegraphics[width=1\linewidth]{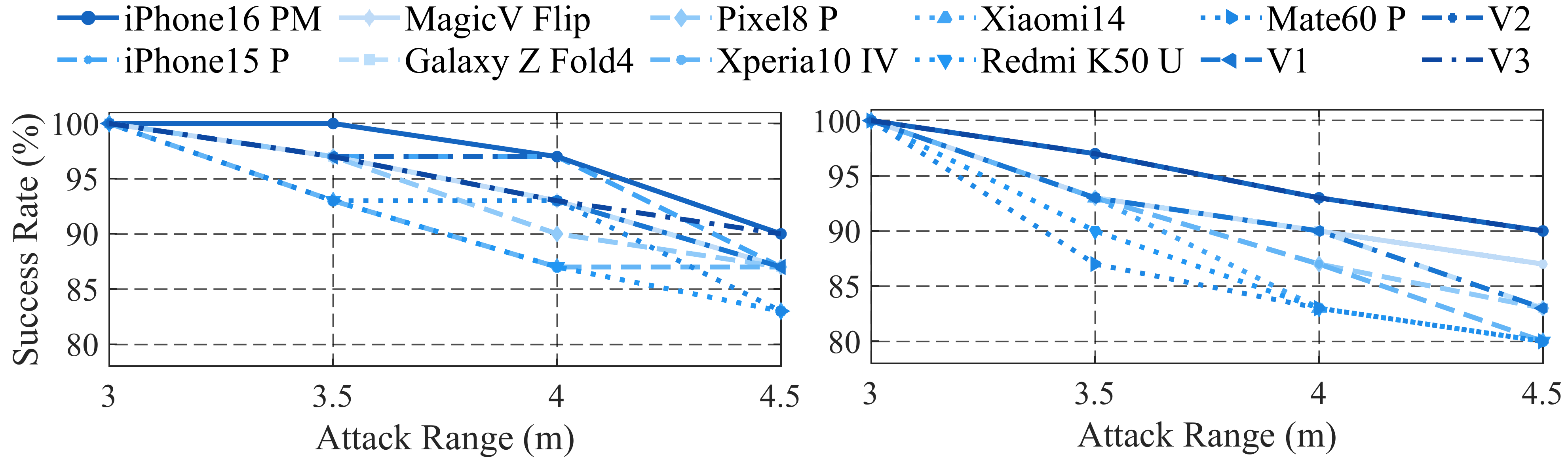}
        \caption{Impact of moving target (left: 2 m/s, right: 3 m/s).}
        \label{moving target}
     \vspace{-10pt}
 \end{figure}




\subsubsection{\SystemName metamaterial design\label{B4}}
To suit outdoor mobile eavesdropping, \SystemName’s metamaterial design balances audio quality and portability. Comparisons with designs that added or removed specific frequency bands show that adding bands yields RSA close to \SystemName, increasing only 0.65\%–1.5\%, while removing bands reduces RSA by 12.3\%–14.5\%, confirming the balance between performance and portability.


\subsection{Performance in Real-world Scenarios \label{sec:prw}}

\subsubsection{Noise impact\label{C1}}
To assess environmental noise, we tested \SystemName under wind (20–300 Hz), thunderstorm (20–200 Hz), traffic (30–300 Hz), and industrial noise (50–500 Hz) (Fig.~\ref{noise}). At 60 dB, average RSA reached 4.05 m, 23.1\% higher than without the algorithm. At 70 dB, RSA without the \textit{Noise Suppression Algorithm} dropped to 0.38 m, while \SystemName maintained 3.53 m, exceeding the \textit{safe distance}.

\subsubsection{Attacks against walking victims\label{C2}}
With its portability and robustness, \SystemName effectively eavesdrops on moving targets. At 4.5 m, average success rates for targets walking at 2 m/s and 3 m/s were 86.7\% and 84.4\%, respectively (Fig.~\ref{moving target}).

\subsubsection{Impact of wind speed}\label{C3}
\begin{figure}[!t]
\centering
\subfloat[]{
		\includegraphics[scale=0.06]{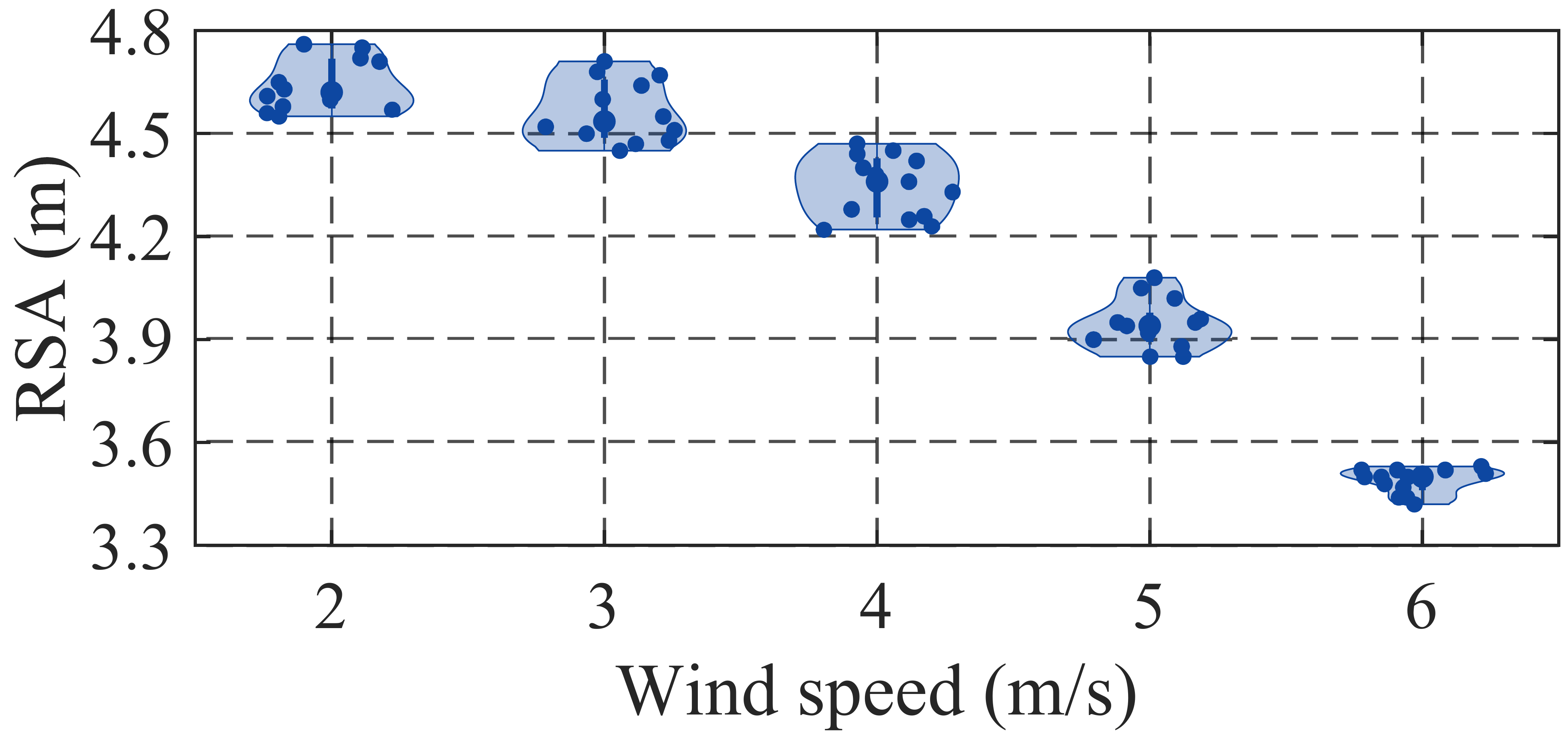}
  \label{WS}}
\hspace{0.1cm}
\subfloat[]{
		\includegraphics[scale=0.06]{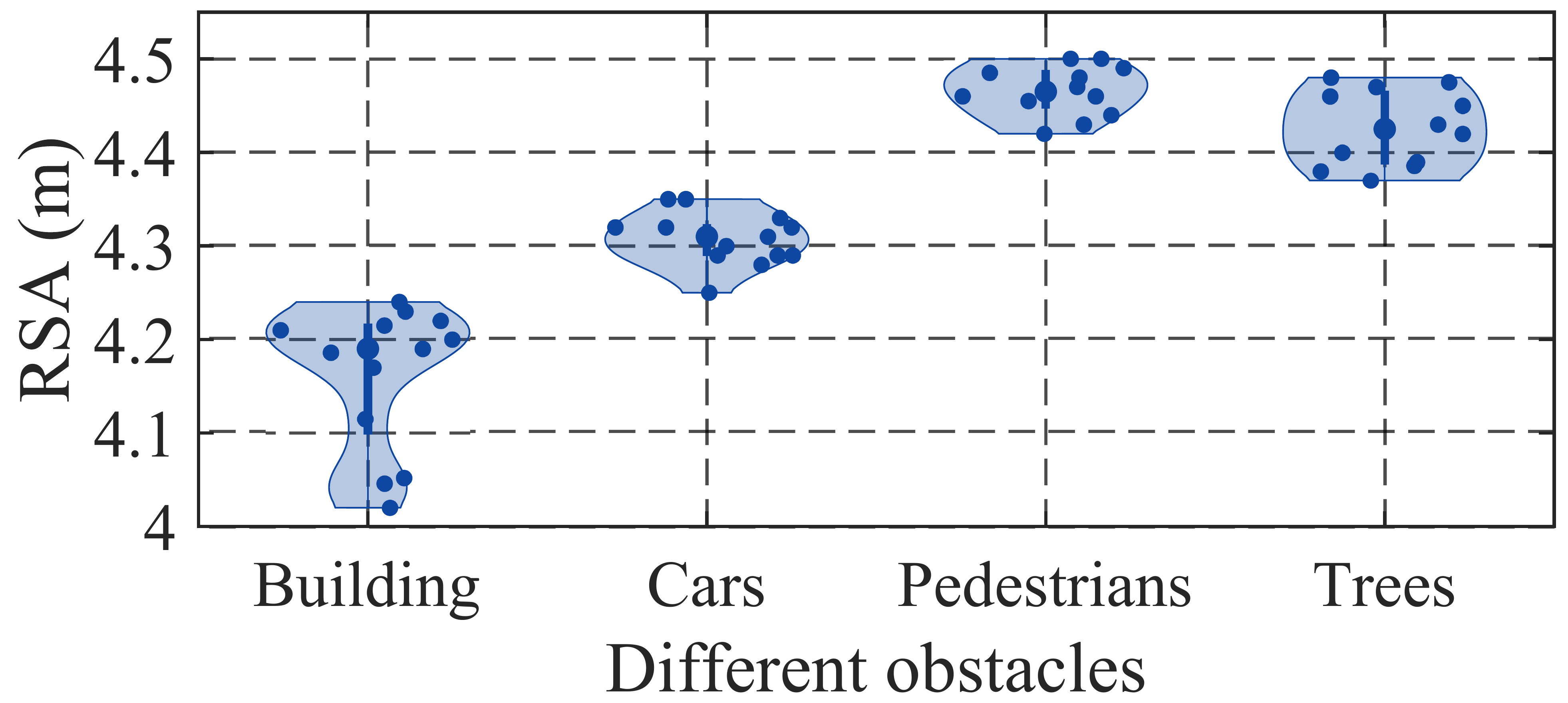}
  \label{OB}}
\caption{(a) Impact of wind. (b) Impact of obstacles.}
\label{2223}
\vspace{-5pt}
\end{figure}

Wind can affect sound propagation, impacting eavesdropping performance. We tested \SystemName under wind speeds of 2–6 m/s; at 6 m/s, the average RSA remained 3.4 m (Fig.~\ref{WS}). This is thanks to the omnidirectional nature of the system’s metamaterials, which capture sound from multiple directions and maintain stable performance.

\subsubsection{Impact of obstacles\label{C4}}
Obstacles partially block sound. Trees, pedestrians, and partially blocked doors/windows were tested. Partially blocked doors/windows reduced RSA from 4.6 m to 4.1 m, but omnidirectional reception allows \SystemName to capture diffracted sound and maintain stable performance.



\subsection{Compared to Prior Methods\label{D}}

We compared \SystemName with prior eavesdropping attacks (Table~\ref{COMPARE}). Only mmSpy~\cite{mmspy}, mmEve~\cite{mmeve}, EarSpy~\cite{earspy}, and Vibphone~\cite{Vibphone} intercept earpiece audio, but EarSpy~\cite{earspy} and Vibphone~\cite{Vibphone} require device infiltration. mmSpy~\cite{mmspy} is limited to 3 m, while mmEve~\cite{mmeve} requires continuous millimeter-wave alignment. Most alternatives rely on costly millimeter-wave radars~\cite{mmspy,mmeve,mmecho} or bulky optical equipment~\cite{lamphone}, often with machine learning overhead. In contrast, \SystemName enables portable, low-cost, non-intrusive eavesdropping suited for outdoor scenarios.

\begin{table}[t!]
    \scriptsize
    \caption{Performance compared to prior work}
    \label{COMPARE}
    \vspace{1mm}
    \centering
    \setlength{\tabcolsep}{3.2pt} 
    \begin{tabular} {p{1.7cm}lllllll}
    \toprule
    \textbf{\makecell[l]{System\\ Name}} & \textbf{\makecell[l]{Earpiece\\ Attack}}  & \textbf{\makecell[l]{Non\\invasive}} & \textbf{\makecell[l]{Over\\ 3 m}} & \textbf{\makecell[l]{No\\ aiming}}&
    \textbf{\makecell[l]{Portable}}&
    \textbf{\makecell[l]{Off-the-\\ shelf}}&
    \textbf{\makecell[l]{User-\\ friendly}}
    \\
    \midrule
    \rowcolor{gray!20}  mmEve \cite{mmeve}& Yes & Yes & Yes & No & Yes & No & No\\
    \rowcolor{gray!20}  mmSpy \cite{mmspy} & Yes & Yes & No & No & Yes & No & No\\
    \rowcolor{gray!20} mmEcho \cite{mmecho}& No & Yes & Yes & No & Yes & No & No\\

    EarSpy \cite{earspy} & Yes & No & Yes & Yes & Yes & Yes & No  \\
    AccelEve \cite{AccelEve} & No & No & Yes & Yes & Yes & Yes & No\\
    Vibphone \cite{Vibphone} & Yes & No & Yes & Yes & Yes & Yes & No\\
    \rowcolor{gray!20}  LidarPhone \cite{LidarPhone}&  No & No & Yes & No & Yes & Yes & No\\
    \rowcolor{gray!20}  Lamphone\cite{lamphone}&  No & Yes & Yes & No & No & No & No \\ 

    \makecell[l]{\textbf{SuperEar}}& \textbf{Yes}& \textbf{Yes}& \textbf{Yes}&\textbf{Yes} & \textbf{Yes}&\textbf{Yes}&\textbf{Yes}\\  
    \bottomrule
    \end{tabular}
      \vspace{-5mm}
\end{table}

\section{DISCUSSION} \label{chap:7}
\cparagraph{Robustness}
\SystemName remains effective under wind (up to 6 m/s), noise (up to 70 dB), and obstacles, though performance may drop in harsher, noisy environments.

\cparagraph{Earphone scenarios}
Current design cannot reliably capture audio from wired or wireless earphones due to weaker leakage; only low-volume speech may be intercepted. This marks the boundary of the current threat model.

\cparagraph{Concealment}
The prototype is compact, fits in a handbag, and works at 4–5 m, typical for public spaces. Concealment is harder in sparsely populated or monitored areas.

\cparagraph{Countermeasures}
Soundproofing (e.g., fiberboards, acoustic foam~\cite{defend1,absorbx,absorby}) or loud white noise can block leakage, but are costly or intrusive outdoors. Practical defenses include portable built-in protections, like directional-transmission earpieces or active-noise-canceling wireless earphones optimized for 250–1000 Hz.

\section{CONCLUSION} \label{chap:9}
We present \SystemName, the first system experimentally demonstrating acoustic metamaterial–based eavesdropping on moving targets outdoors. Unlike prior non‑intrusive methods, it reliably captures phone-call audio at several meters. Key advances—combining multiple metamaterials for low-frequency gain, optimizing structures for portability, and applying filtering to reduce noise and artifacts—enable effective audio reconstruction in realistic environments. Our evaluation shows \SystemName outperforms existing attacks, revealing a new privacy threat and highlighting the need for improved defenses against acoustic side channels.

\section{Ethical Considerations} \label{chap:13}
We present a novel acoustic eavesdropping attack using acoustic metamaterials, framed with explicit ethical considerations. Our work highlights overlooked privacy risks in outdoor phone calls and motivates countermeasure development. To reduce misuse, we focus on feasibility and limitations rather than replication details, aiming to guide future defenses and policy.

\bibliographystyle{ACM-Reference-Format}
\balance
\bibliography{references}


\begin{thebibliography}{61}


\ifx \showCODEN    \undefined \def \showCODEN     #1{\unskip}     \fi
\ifx \showISBNx    \undefined \def \showISBNx     #1{\unskip}     \fi
\ifx \showISBNxiii \undefined \def \showISBNxiii  #1{\unskip}     \fi
\ifx \showISSN     \undefined \def \showISSN      #1{\unskip}     \fi
\ifx \showLCCN     \undefined \def \showLCCN      #1{\unskip}     \fi
\ifx \shownote     \undefined \def \shownote      #1{#1}          \fi
\ifx \showarticletitle \undefined \def \showarticletitle #1{#1}   \fi
\ifx \showURL      \undefined \def \showURL       {\relax}        \fi
\providecommand\bibfield[2]{#2}
\providecommand\bibinfo[2]{#2}
\providecommand\natexlab[1]{#1}
\providecommand\showeprint[2][]{arXiv:#2}

\bibitem[Goo(text)]%
        {Google}
 \bibinfo{year}{\url{https://cloud.google.com/speech-to-text}}\natexlab{}.
\newblock \bibinfo{title}{Google Text-to-Speech AI}.
\newblock
\newblock
\shownote{Last accessed: 2024-8-6}.


\bibitem[Hua(ones)]%
        {Huawei}
 \bibinfo{year}{\url{https://consumer.huawei.com/cn/phones/}}\natexlab{}.
\newblock \bibinfo{title}{Huawei}.
\newblock
\newblock
\shownote{Last accessed: 2024-10-14}.


\bibitem[Sox(ssox)]%
        {Sox}
 \bibinfo{year}{\url{https://sourceforge.net/projects/sox/}}\natexlab{}.
\newblock \bibinfo{title}{Sox}.
\newblock
\newblock
\shownote{Last accessed: 2024-10-6}.


\bibitem[App(hone)]%
        {Apple}
 \bibinfo{year}{\url{https://www.apple.com.cn/iphone/}}\natexlab{}.
\newblock \bibinfo{title}{Apple}.
\newblock
\newblock
\shownote{Last accessed: 2024-10-14}.


\bibitem[HON(ones)]%
        {HONOR}
 \bibinfo{year}{\url{https://www.honor.com/cn/phones/}}\natexlab{}.
\newblock \bibinfo{title}{HONOR}.
\newblock
\newblock
\shownote{Last accessed: 2024-10-14}.


\bibitem[Xia(icom)]%
        {Xiaomi}
 \bibinfo{year}{\url{https://www.mi.com/}}\natexlab{}.
\newblock \bibinfo{title}{Xiaomi}.
\newblock
\newblock
\shownote{Last accessed: 2024-10-14}.


\bibitem[Sam(ones)]%
        {Samsung}
 \bibinfo{year}{\url{https://www.samsung.com.cn/smartphones/all-smartphones/}}\natexlab{}.
\newblock \bibinfo{title}{Samsung}.
\newblock
\newblock
\shownote{Last accessed: 2024-10-14}.


\bibitem[Son(m1gk)]%
        {Sony}
 \bibinfo{year}{\url{https://www.sony.com.hk/zh/smartphones?srsltid=AfmBOopWInT4KRcdpk37iCnv3688Jc44KDiGgdUlSnzyn-PZlq-Mm1gk}}\natexlab{}.
\newblock \bibinfo{title}{Sony}.
\newblock
\newblock
\shownote{Last accessed: 2024-10-14}.


\bibitem[Anand and Saxena(2018)]%
        {speechless}
\bibfield{author}{\bibinfo{person}{S~Abhishek Anand} {and} \bibinfo{person}{Nitesh Saxena}.} \bibinfo{year}{2018}\natexlab{}.
\newblock \showarticletitle{Speechless: Analyzing the threat to speech privacy from smartphone motion sensors}. In \bibinfo{booktitle}{\emph{2018 IEEE Symposium on Security and Privacy (SP)}}. IEEE, \bibinfo{pages}{1000--1017}.
\newblock


\bibitem[Assouar et~al\mbox{.}(2018)]%
        {r11}
\bibfield{author}{\bibinfo{person}{Badreddine Assouar}, \bibinfo{person}{Bin Liang}, \bibinfo{person}{Ying Wu}, \bibinfo{person}{Yong Li}, \bibinfo{person}{Jian-Chun Cheng}, {and} \bibinfo{person}{Yun Jing}.} \bibinfo{year}{2018}\natexlab{}.
\newblock \showarticletitle{Acoustic metasurfaces}.
\newblock \bibinfo{journal}{\emph{Nature Reviews Materials}} \bibinfo{volume}{3}, \bibinfo{number}{12} (\bibinfo{year}{2018}), \bibinfo{pages}{460--472}.
\newblock


\bibitem[Ba et~al\mbox{.}(2007)]%
        {MVDR}
\bibfield{author}{\bibinfo{person}{Demba~E Ba}, \bibinfo{person}{Dinei Flor{\^e}ncio}, {and} \bibinfo{person}{Cha Zhang}.} \bibinfo{year}{2007}\natexlab{}.
\newblock \showarticletitle{Enhanced MVDR beamforming for arrays of directional microphones}. In \bibinfo{booktitle}{\emph{2007 IEEE international conference on multimedia and expo}}. IEEE, \bibinfo{pages}{1307--1310}.
\newblock


\bibitem[Ba et~al\mbox{.}(2020)]%
        {AccelEve}
\bibfield{author}{\bibinfo{person}{Zhongjie Ba}, \bibinfo{person}{Tianhang Zheng}, \bibinfo{person}{Xinyu Zhang}, \bibinfo{person}{Zhan Qin}, \bibinfo{person}{Baochun Li}, \bibinfo{person}{Xue Liu}, {and} \bibinfo{person}{Kui Ren}.} \bibinfo{year}{2020}\natexlab{}.
\newblock \showarticletitle{Learning-based Practical Smartphone Eavesdropping with Built-in Accelerometer.}. In \bibinfo{booktitle}{\emph{NDSS}}, Vol.~\bibinfo{volume}{2020}. \bibinfo{pages}{1--18}.
\newblock


\bibitem[Basak and Gowda(2022)]%
        {mmspy}
\bibfield{author}{\bibinfo{person}{Suryoday Basak} {and} \bibinfo{person}{Mahanth Gowda}.} \bibinfo{year}{2022}\natexlab{}.
\newblock \showarticletitle{mmspy: Spying phone calls using mmwave radars}. In \bibinfo{booktitle}{\emph{2022 IEEE Symposium on Security and Privacy (SP)}}. IEEE, \bibinfo{pages}{1211--1228}.
\newblock


\bibitem[Catford(2001)]%
        {catford2001practical}
\bibfield{author}{\bibinfo{person}{John~Cunnison Catford}.} \bibinfo{year}{2001}\natexlab{}.
\newblock \bibinfo{booktitle}{\emph{A practical introduction to phonetics}}.
\newblock \bibinfo{publisher}{Oxford University Press}.
\newblock


\bibitem[Cheng et~al\mbox{.}(2015)]%
        {ultra}
\bibfield{author}{\bibinfo{person}{Y Cheng}, \bibinfo{person}{C Zhou}, \bibinfo{person}{BG Yuan}, \bibinfo{person}{DJ Wu}, \bibinfo{person}{Q Wei}, {and} \bibinfo{person}{XJ Liu}.} \bibinfo{year}{2015}\natexlab{}.
\newblock \showarticletitle{Ultra-sparse metasurface for high reflection of low-frequency sound based on artificial Mie resonances}.
\newblock \bibinfo{journal}{\emph{Nature materials}} \bibinfo{volume}{14}, \bibinfo{number}{10} (\bibinfo{year}{2015}), \bibinfo{pages}{1013--1019}.
\newblock


\bibitem[Davis et~al\mbox{.}(2014)]%
        {VisualMicrophone}
\bibfield{author}{\bibinfo{person}{Abe Davis}, \bibinfo{person}{Michael Rubinstein}, \bibinfo{person}{Neal Wadhwa}, \bibinfo{person}{Gautham~J Mysore}, \bibinfo{person}{Fredo Durand}, {and} \bibinfo{person}{William~T Freeman}.} \bibinfo{year}{2014}\natexlab{}.
\newblock \showarticletitle{The visual microphone: Passive recovery of sound from video}.
\newblock \bibinfo{journal}{\emph{ACM Transactions on Graphics}} (\bibinfo{year}{2014}).
\newblock


\bibitem[Deng et~al\mbox{.}(2013)]%
        {MVDR2}
\bibfield{author}{\bibinfo{person}{Aidong Deng}, \bibinfo{person}{Hang Tong}, \bibinfo{person}{Jianeng Tang}, \bibinfo{person}{Hao Cao}, \bibinfo{person}{Kang Qin}, {and} \bibinfo{person}{Xi Yan}.} \bibinfo{year}{2013}\natexlab{}.
\newblock \showarticletitle{Study on location algorithms of beamforming based on MVDR}.
\newblock \bibinfo{journal}{\emph{Applied Mathematics \& Information Sciences}} \bibinfo{volume}{7}, \bibinfo{number}{6} (\bibinfo{year}{2013}), \bibinfo{pages}{2455}.
\newblock


\bibitem[Hafter et~al\mbox{.}(2008)]%
        {selective2}
\bibfield{author}{\bibinfo{person}{Ervin~R Hafter}, \bibinfo{person}{Anastasios Sarampalis}, {and} \bibinfo{person}{Psyche Loui}.} \bibinfo{year}{2008}\natexlab{}.
\newblock \showarticletitle{Auditory attention and filters}.
\newblock In \bibinfo{booktitle}{\emph{Auditory perception of sound sources}}. \bibinfo{publisher}{Springer}, \bibinfo{pages}{115--142}.
\newblock


\bibitem[Han et~al\mbox{.}(2017)]%
        {PitchIn}
\bibfield{author}{\bibinfo{person}{Jun Han}, \bibinfo{person}{Albert~Jin Chung}, {and} \bibinfo{person}{Patrick Tague}.} \bibinfo{year}{2017}\natexlab{}.
\newblock \showarticletitle{Pitchln: eavesdropping via intelligible speech reconstruction using non-acoustic sensor fusion}. In \bibinfo{booktitle}{\emph{Proceedings of the 16th ACM/IEEE International Conference on Information Processing in Sensor Networks}}. \bibinfo{pages}{181--192}.
\newblock


\bibitem[Heintze et~al\mbox{.}(2008)]%
        {500noiserange1}
\bibfield{author}{\bibinfo{person}{Olaf Heintze}, \bibinfo{person}{Volker Wittstock}, {and} \bibinfo{person}{Carl~Fredrik Hartung}.} \bibinfo{year}{2008}\natexlab{}.
\newblock \showarticletitle{Sound Radiation of a Large Truck Oil Pan: Estimation and Experimental Investigation}.
\newblock \bibinfo{journal}{\emph{Journal of the Acoustical Society of America}} \bibinfo{volume}{123}, \bibinfo{number}{5} (\bibinfo{year}{2008}), \bibinfo{pages}{3171--3171}.
\newblock


\bibitem[Hu et~al\mbox{.}(2023)]%
        {mmecho}
\bibfield{author}{\bibinfo{person}{Pengfei Hu}, \bibinfo{person}{Wenhao Li}, \bibinfo{person}{Riccardo Spolaor}, {and} \bibinfo{person}{Xiuzhen Cheng}.} \bibinfo{year}{2023}\natexlab{}.
\newblock \showarticletitle{mmecho: A mmwave-based acoustic eavesdropping method}. In \bibinfo{booktitle}{\emph{Proceedings of the ACM Turing Award Celebration Conference-China 2023}}. \bibinfo{pages}{138--140}.
\newblock


\bibitem[Hu et~al\mbox{.}(2022a)]%
        {milliear}
\bibfield{author}{\bibinfo{person}{Pengfei Hu}, \bibinfo{person}{Yifan Ma}, \bibinfo{person}{Panneer~Selvam Santhalingam}, \bibinfo{person}{Parth~H Pathak}, {and} \bibinfo{person}{Xiuzhen Cheng}.} \bibinfo{year}{2022}\natexlab{a}.
\newblock \showarticletitle{Milliear: Millimeter-wave acoustic eavesdropping with unconstrained vocabulary}. In \bibinfo{booktitle}{\emph{IEEE INFOCOM 2022-IEEE Conference on Computer Communications}}. IEEE, \bibinfo{pages}{11--20}.
\newblock


\bibitem[Hu et~al\mbox{.}(2022b)]%
        {accear}
\bibfield{author}{\bibinfo{person}{Pengfei Hu}, \bibinfo{person}{Hui Zhuang}, \bibinfo{person}{Panneer~Selvam Santhalingam}, \bibinfo{person}{Riccardo Spolaor}, \bibinfo{person}{Parth Pathak}, \bibinfo{person}{Guoming Zhang}, {and} \bibinfo{person}{Xiuzhen Cheng}.} \bibinfo{year}{2022}\natexlab{b}.
\newblock \showarticletitle{Accear: Accelerometer acoustic eavesdropping with unconstrained vocabulary}. In \bibinfo{booktitle}{\emph{2022 IEEE Symposium on Security and Privacy (SP)}}. IEEE, \bibinfo{pages}{1757--1773}.
\newblock


\bibitem[Ji et~al\mbox{.}(2023)]%
        {focus4}
\bibfield{author}{\bibinfo{person}{Jun Ji}, \bibinfo{person}{Chuming Zhao}, \bibinfo{person}{Frank Yao}, \bibinfo{person}{Tetsuro Oishi}, \bibinfo{person}{John Stewart}, {and} \bibinfo{person}{Yun Jing}.} \bibinfo{year}{2023}\natexlab{}.
\newblock \showarticletitle{Metamaterial-Augmented Head-Mounted Audio Module}.
\newblock \bibinfo{journal}{\emph{Advanced Materials Technologies}} \bibinfo{volume}{8}, \bibinfo{number}{19} (\bibinfo{year}{2023}), \bibinfo{pages}{2300834}.
\newblock


\bibitem[Katinas et~al\mbox{.}(2016)]%
        {300noiserange1}
\bibfield{author}{\bibinfo{person}{Vladislovas Katinas}, \bibinfo{person}{Mantas Mar{\v{c}}iukaitis}, {and} \bibinfo{person}{Marijona Tama{\v{s}}auskien{\.e}}.} \bibinfo{year}{2016}\natexlab{}.
\newblock \showarticletitle{Analysis of the wind turbine noise emissions and impact on the environment}.
\newblock \bibinfo{journal}{\emph{Renewable and Sustainable Energy Reviews}}  \bibinfo{volume}{58} (\bibinfo{year}{2016}), \bibinfo{pages}{825--831}.
\newblock


\bibitem[Kominek et~al\mbox{.}(2008)]%
        {Mel}
\bibfield{author}{\bibinfo{person}{John Kominek}, \bibinfo{person}{Tanja Schultz}, {and} \bibinfo{person}{Alan~W Black}.} \bibinfo{year}{2008}\natexlab{}.
\newblock \showarticletitle{Synthesizer voice quality of new languages calibrated with mean mel cepstral distortion.}. In \bibinfo{booktitle}{\emph{SLTU}}. \bibinfo{pages}{63--68}.
\newblock


\bibitem[Kwong et~al\mbox{.}(2019)]%
        {HDD}
\bibfield{author}{\bibinfo{person}{Andrew Kwong}, \bibinfo{person}{Wenyuan Xu}, {and} \bibinfo{person}{Kevin Fu}.} \bibinfo{year}{2019}\natexlab{}.
\newblock \showarticletitle{Hard drive of hearing: Disks that eavesdrop with a synthesized microphone}. In \bibinfo{booktitle}{\emph{2019 IEEE symposium on security and privacy (SP)}}. IEEE, \bibinfo{pages}{905--919}.
\newblock


\bibitem[Lei et~al\mbox{.}(2023)]%
        {MieResonances2}
\bibfield{author}{\bibinfo{person}{Yunzhong Lei}, \bibinfo{person}{Jiu~Hui Wu}, \bibinfo{person}{Libo Wang}, \bibinfo{person}{Yao Huang}, {and} \bibinfo{person}{Jiamin Niu}.} \bibinfo{year}{2023}\natexlab{}.
\newblock \showarticletitle{Deep sub-wavelength acoustic transmission enhancement and whisper via the monopole resonance in meta-cavities}.
\newblock \bibinfo{journal}{\emph{Applied Acoustics}}  \bibinfo{volume}{203} (\bibinfo{year}{2023}), \bibinfo{pages}{109227}.
\newblock


\bibitem[Luck(2016)]%
        {selective3}
\bibfield{author}{\bibinfo{person}{Steven~J Luck}.} \bibinfo{year}{2016}\natexlab{}.
\newblock \showarticletitle{Neurophysiology of selective attention}.
\newblock In \bibinfo{booktitle}{\emph{Attention}}. \bibinfo{publisher}{Psychology Press}, \bibinfo{pages}{257--295}.
\newblock


\bibitem[Ma and Sheng(2016)]%
        {r12}
\bibfield{author}{\bibinfo{person}{Guancong Ma} {and} \bibinfo{person}{Ping Sheng}.} \bibinfo{year}{2016}\natexlab{}.
\newblock \showarticletitle{Acoustic metamaterials: From local resonances to broad horizons}.
\newblock \bibinfo{journal}{\emph{Science advances}} \bibinfo{volume}{2}, \bibinfo{number}{2} (\bibinfo{year}{2016}), \bibinfo{pages}{e1501595}.
\newblock


\bibitem[Mahdad et~al\mbox{.}(2022)]%
        {earspy}
\bibfield{author}{\bibinfo{person}{Ahmed~Tanvir Mahdad}, \bibinfo{person}{Cong Shi}, \bibinfo{person}{Zhengkun Ye}, \bibinfo{person}{Tianming Zhao}, \bibinfo{person}{Yan Wang}, \bibinfo{person}{Yingying Chen}, {and} \bibinfo{person}{Nitesh Saxena}.} \bibinfo{year}{2022}\natexlab{}.
\newblock \showarticletitle{Earspy: Spying caller speech and identity through tiny vibrations of smartphone ear speakers}.
\newblock \bibinfo{journal}{\emph{arXiv preprint arXiv:2212.12151}} (\bibinfo{year}{2022}).
\newblock


\bibitem[Maruri et~al\mbox{.}(2018)]%
        {V-Speech}
\bibfield{author}{\bibinfo{person}{H{\'e}ctor A~Cordourier Maruri}, \bibinfo{person}{Paulo Lopez-Meyer}, \bibinfo{person}{Jonathan Huang}, \bibinfo{person}{Willem~Marco Beltman}, \bibinfo{person}{Lama Nachman}, {and} \bibinfo{person}{Hong Lu}.} \bibinfo{year}{2018}\natexlab{}.
\newblock \showarticletitle{V-Speech: noise-robust speech capturing glasses using vibration sensors}.
\newblock \bibinfo{journal}{\emph{Proceedings of the ACM on Interactive, Mobile, Wearable and Ubiquitous Technologies}} \bibinfo{volume}{2}, \bibinfo{number}{4} (\bibinfo{year}{2018}), \bibinfo{pages}{1--23}.
\newblock


\bibitem[Michalevsky et~al\mbox{.}(2014)]%
        {gyrophone}
\bibfield{author}{\bibinfo{person}{Yan Michalevsky}, \bibinfo{person}{Dan Boneh}, {and} \bibinfo{person}{Gabi Nakibly}.} \bibinfo{year}{2014}\natexlab{}.
\newblock \showarticletitle{Gyrophone: Recognizing speech from gyroscope signals}. In \bibinfo{booktitle}{\emph{23rd USENIX Security Symposium (USENIX Security 14)}}. \bibinfo{pages}{1053--1067}.
\newblock


\bibitem[Nassi et~al\mbox{.}(2022)]%
        {lamphone}
\bibfield{author}{\bibinfo{person}{Ben Nassi}, \bibinfo{person}{Yaron Pirutin}, \bibinfo{person}{Raz Swisa}, \bibinfo{person}{Adi Shamir}, \bibinfo{person}{Yuval Elovici}, {and} \bibinfo{person}{Boris Zadov}.} \bibinfo{year}{2022}\natexlab{}.
\newblock \showarticletitle{Lamphone: Passive sound recovery from a desk lamp's light bulb vibrations}. In \bibinfo{booktitle}{\emph{31st USENIX Security Symposium (USENIX Security 22)}}. \bibinfo{pages}{4401--4417}.
\newblock


\bibitem[Nechita and N{\u{a}}stac(2018)]%
        {absorby}
\bibfield{author}{\bibinfo{person}{P Nechita} {and} \bibinfo{person}{S N{\u{a}}stac}.} \bibinfo{year}{2018}\natexlab{}.
\newblock \showarticletitle{Foam-formed cellulose composite materials with potential applications in sound insulation}.
\newblock \bibinfo{journal}{\emph{Journal of composite materials}} \bibinfo{volume}{52}, \bibinfo{number}{6} (\bibinfo{year}{2018}), \bibinfo{pages}{747--754}.
\newblock


\bibitem[Peng(2017a)]%
        {defend1}
\bibfield{author}{\bibinfo{person}{L Peng}.} \bibinfo{year}{2017}\natexlab{a}.
\newblock \showarticletitle{Sound absorption and insulation functional composites}.
\newblock In \bibinfo{booktitle}{\emph{Advanced high strength natural fibre composites in construction}}. \bibinfo{publisher}{Elsevier}, \bibinfo{pages}{333--373}.
\newblock


\bibitem[Peng(2017b)]%
        {absorbx}
\bibfield{author}{\bibinfo{person}{L Peng}.} \bibinfo{year}{2017}\natexlab{b}.
\newblock \showarticletitle{Sound absorption and insulation functional composites}.
\newblock In \bibinfo{booktitle}{\emph{Advanced high strength natural fibre composites in construction}}. \bibinfo{publisher}{Elsevier}, \bibinfo{pages}{333--373}.
\newblock


\bibitem[Qu et~al\mbox{.}(2023)]%
        {focus5}
\bibfield{author}{\bibinfo{person}{Sichao Qu}, \bibinfo{person}{Min Yang}, \bibinfo{person}{Yunfei Xu}, \bibinfo{person}{Songwen Xiao}, {and} \bibinfo{person}{Nicholas~X Fang}.} \bibinfo{year}{2023}\natexlab{}.
\newblock \showarticletitle{Reverberation time control by acoustic metamaterials in a small room}.
\newblock \bibinfo{journal}{\emph{Building and Environment}}  \bibinfo{volume}{244} (\bibinfo{year}{2023}), \bibinfo{pages}{110753}.
\newblock


\bibitem[Ren et~al\mbox{.}(2022)]%
        {r15}
\bibfield{author}{\bibinfo{person}{Zhiwen Ren}, \bibinfo{person}{Yuehang Cheng}, \bibinfo{person}{Mingji Chen}, \bibinfo{person}{Xujin Yuan}, {and} \bibinfo{person}{Daining Fang}.} \bibinfo{year}{2022}\natexlab{}.
\newblock \showarticletitle{A compact multifunctional metastructure for Low-frequency broadband sound absorption and crash energy dissipation}.
\newblock \bibinfo{journal}{\emph{Materials \& Design}}  \bibinfo{volume}{215} (\bibinfo{year}{2022}), \bibinfo{pages}{110462}.
\newblock


\bibitem[Richmond(2008)]%
        {r99}
\bibfield{author}{\bibinfo{person}{Virginia~P Richmond}.} \bibinfo{year}{2008}\natexlab{}.
\newblock \showarticletitle{Nonverbal behavior in interpersonal relations}.
\newblock \bibinfo{journal}{\emph{(No Title)}} (\bibinfo{year}{2008}), \bibinfo{pages}{366}.
\newblock


\bibitem[Roy and Roy~Choudhury(2016)]%
        {VibraPhone}
\bibfield{author}{\bibinfo{person}{Nirupam Roy} {and} \bibinfo{person}{Romit Roy~Choudhury}.} \bibinfo{year}{2016}\natexlab{}.
\newblock \showarticletitle{Listening through a vibration motor}. In \bibinfo{booktitle}{\emph{Proceedings of the 14th Annual International Conference on Mobile Systems, Applications, and Services}}. \bibinfo{pages}{57--69}.
\newblock


\bibitem[Sami et~al\mbox{.}(2020)]%
        {LidarPhone}
\bibfield{author}{\bibinfo{person}{Sriram Sami}, \bibinfo{person}{Yimin Dai}, \bibinfo{person}{Sean Rui~Xiang Tan}, \bibinfo{person}{Nirupam Roy}, {and} \bibinfo{person}{Jun Han}.} \bibinfo{year}{2020}\natexlab{}.
\newblock \showarticletitle{Spying with your robot vacuum cleaner: eavesdropping via lidar sensors}. In \bibinfo{booktitle}{\emph{Proceedings of the 18th Conference on Embedded Networked Sensor Systems}}. \bibinfo{pages}{354--367}.
\newblock


\bibitem[Su et~al\mbox{.}(2021)]%
        {Vibphone}
\bibfield{author}{\bibinfo{person}{Weigao Su}, \bibinfo{person}{Daibo Liu}, \bibinfo{person}{Taiyuan Zhang}, {and} \bibinfo{person}{Hongbo Jiang}.} \bibinfo{year}{2021}\natexlab{}.
\newblock \showarticletitle{Towards device independent eavesdropping on telephone conversations with built-in accelerometer}.
\newblock \bibinfo{journal}{\emph{Proceedings of the ACM on Interactive, Mobile, Wearable and Ubiquitous Technologies}} \bibinfo{volume}{5}, \bibinfo{number}{4} (\bibinfo{year}{2021}), \bibinfo{pages}{1--29}.
\newblock


\bibitem[Sun et~al\mbox{.}(2019)]%
        {MieResonances}
\bibfield{author}{\bibinfo{person}{Ye-Yang Sun}, \bibinfo{person}{Jian-Ping Xia}, \bibinfo{person}{Hong-Xiang Sun}, \bibinfo{person}{Shou-Qi Yuan}, \bibinfo{person}{Yong Ge}, {and} \bibinfo{person}{Xiao-Jun Liu}.} \bibinfo{year}{2019}\natexlab{}.
\newblock \showarticletitle{Dual-Band Fano Resonance of Low-Frequency Sound Based on Artificial Mie Resonances}.
\newblock \bibinfo{journal}{\emph{Advanced Science}} \bibinfo{volume}{6}, \bibinfo{number}{20} (\bibinfo{year}{2019}), \bibinfo{pages}{1901307}.
\newblock


\bibitem[Sundstrom and Altman(1976)]%
        {privacy2}
\bibfield{author}{\bibinfo{person}{Eric Sundstrom} {and} \bibinfo{person}{Irwin Altman}.} \bibinfo{year}{1976}\natexlab{}.
\newblock \showarticletitle{Interpersonal relationships and personal space: Research review and theoretical model}.
\newblock \bibinfo{journal}{\emph{Human Ecology}}  \bibinfo{volume}{4} (\bibinfo{year}{1976}), \bibinfo{pages}{47--67}.
\newblock


\bibitem[Walsh et~al\mbox{.}(2014)]%
        {selective1}
\bibfield{author}{\bibinfo{person}{Kyle~P Walsh}, \bibinfo{person}{Edward~G Pasanen}, {and} \bibinfo{person}{Dennis McFadden}.} \bibinfo{year}{2014}\natexlab{}.
\newblock \showarticletitle{Selective attention reduces physiological noise in the external ear canals of humans. I: Auditory attention}.
\newblock \bibinfo{journal}{\emph{Hearing research}}  \bibinfo{volume}{312} (\bibinfo{year}{2014}), \bibinfo{pages}{143--159}.
\newblock


\bibitem[Wang et~al\mbox{.}(2022)]%
        {mmeve}
\bibfield{author}{\bibinfo{person}{Chao Wang}, \bibinfo{person}{Feng Lin}, \bibinfo{person}{Tiantian Liu}, \bibinfo{person}{Kaidi Zheng}, \bibinfo{person}{Zhibo Wang}, \bibinfo{person}{Zhengxiong Li}, \bibinfo{person}{Ming-Chun Huang}, \bibinfo{person}{Wenyao Xu}, {and} \bibinfo{person}{Kui Ren}.} \bibinfo{year}{2022}\natexlab{}.
\newblock \showarticletitle{mmEve: eavesdropping on smartphone's earpiece via COTS mmWave device}. In \bibinfo{booktitle}{\emph{Proceedings of the 28th Annual International Conference on Mobile Computing And Networking}}. \bibinfo{pages}{338--351}.
\newblock


\bibitem[Wang et~al\mbox{.}(2021)]%
        {Tag-Bug}
\bibfield{author}{\bibinfo{person}{Chuyu Wang}, \bibinfo{person}{Lei Xie}, \bibinfo{person}{Yuancan Lin}, \bibinfo{person}{Wei Wang}, \bibinfo{person}{Yingying Chen}, \bibinfo{person}{Yanling Bu}, \bibinfo{person}{Kai Zhang}, {and} \bibinfo{person}{Sanglu Lu}.} \bibinfo{year}{2021}\natexlab{}.
\newblock \showarticletitle{Thru-the-wall eavesdropping on loudspeakers via RFID by capturing sub-mm level vibration}.
\newblock \bibinfo{journal}{\emph{Proceedings of the ACM on Interactive, Mobile, Wearable and Ubiquitous Technologies}} \bibinfo{volume}{5}, \bibinfo{number}{4} (\bibinfo{year}{2021}), \bibinfo{pages}{1--25}.
\newblock


\bibitem[Wang et~al\mbox{.}(2014)]%
        {WiHear}
\bibfield{author}{\bibinfo{person}{Guanhua Wang}, \bibinfo{person}{Yongpan Zou}, \bibinfo{person}{Zimu Zhou}, \bibinfo{person}{Kaishun Wu}, {and} \bibinfo{person}{Lionel~M Ni}.} \bibinfo{year}{2014}\natexlab{}.
\newblock \showarticletitle{We can hear you with Wi-Fi!}. In \bibinfo{booktitle}{\emph{Proceedings of the 20th annual international conference on Mobile computing and networking}}. \bibinfo{pages}{593--604}.
\newblock


\bibitem[Wang et~al\mbox{.}(2023)]%
        {voicelistener}
\bibfield{author}{\bibinfo{person}{Lei Wang}, \bibinfo{person}{Meng Chen}, \bibinfo{person}{Li Lu}, \bibinfo{person}{Zhongjie Ba}, \bibinfo{person}{Feng Lin}, {and} \bibinfo{person}{Kui Ren}.} \bibinfo{year}{2023}\natexlab{}.
\newblock \showarticletitle{Voicelistener: A training-free and universal eavesdropping attack on built-in speakers of mobile devices}.
\newblock \bibinfo{journal}{\emph{Proceedings of the ACM on Interactive, Mobile, Wearable and Ubiquitous Technologies}} \bibinfo{volume}{7}, \bibinfo{number}{1} (\bibinfo{year}{2023}), \bibinfo{pages}{1--22}.
\newblock


\bibitem[Wang et~al\mbox{.}(2024)]%
        {300noiserange2}
\bibfield{author}{\bibinfo{person}{Wenjie Wang}, \bibinfo{person}{Yan Yan}, \bibinfo{person}{Yongnian Zhao}, {and} \bibinfo{person}{Yu Xue}.} \bibinfo{year}{2024}\natexlab{}.
\newblock \showarticletitle{Studies on the Experimental Measurement of the Low-Frequency Aerodynamic Noise of Large Wind Turbines}.
\newblock \bibinfo{journal}{\emph{Energies}} \bibinfo{volume}{17}, \bibinfo{number}{7} (\bibinfo{year}{2024}), \bibinfo{pages}{1609}.
\newblock


\bibitem[Wang et~al\mbox{.}(2020)]%
        {uwhear}
\bibfield{author}{\bibinfo{person}{Ziqi Wang}, \bibinfo{person}{Zhe Chen}, \bibinfo{person}{Akash~Deep Singh}, \bibinfo{person}{Luis Garcia}, \bibinfo{person}{Jun Luo}, {and} \bibinfo{person}{Mani~B Srivastava}.} \bibinfo{year}{2020}\natexlab{}.
\newblock \showarticletitle{UWHear: Through-wall extraction and separation of audio vibrations using wireless signals}. In \bibinfo{booktitle}{\emph{Proceedings of the 18th Conference on Embedded Networked Sensor Systems}}. \bibinfo{pages}{1--14}.
\newblock


\bibitem[Wei et~al\mbox{.}(2015)]%
        {ART}
\bibfield{author}{\bibinfo{person}{Teng Wei}, \bibinfo{person}{Shu Wang}, \bibinfo{person}{Anfu Zhou}, {and} \bibinfo{person}{Xinyu Zhang}.} \bibinfo{year}{2015}\natexlab{}.
\newblock \showarticletitle{Acoustic eavesdropping through wireless vibrometry}. In \bibinfo{booktitle}{\emph{Proceedings of the 21st Annual International Conference on Mobile Computing and Networking}}. \bibinfo{pages}{130--141}.
\newblock


\bibitem[Wieser et~al\mbox{.}(2010)]%
        {privacy3}
\bibfield{author}{\bibinfo{person}{Matthias~J Wieser}, \bibinfo{person}{Paul Pauli}, \bibinfo{person}{Miriam Grosseibl}, \bibinfo{person}{Ina Molzow}, {and} \bibinfo{person}{Andreas M{\"u}hlberger}.} \bibinfo{year}{2010}\natexlab{}.
\newblock \showarticletitle{Virtual social interactions in social anxiety—the impact of sex, gaze, and interpersonal distance}.
\newblock \bibinfo{journal}{\emph{Cyberpsychology, Behavior, and Social Networking}} \bibinfo{volume}{13}, \bibinfo{number}{5} (\bibinfo{year}{2010}), \bibinfo{pages}{547--554}.
\newblock


\bibitem[Xu et~al\mbox{.}(2019)]%
        {waveear}
\bibfield{author}{\bibinfo{person}{Chenhan Xu}, \bibinfo{person}{Zhengxiong Li}, \bibinfo{person}{Hanbin Zhang}, \bibinfo{person}{Aditya~Singh Rathore}, \bibinfo{person}{Huining Li}, \bibinfo{person}{Chen Song}, \bibinfo{person}{Kun Wang}, {and} \bibinfo{person}{Wenyao Xu}.} \bibinfo{year}{2019}\natexlab{}.
\newblock \showarticletitle{Waveear: Exploring a mmwave-based noise-resistant speech sensing for voice-user interface}. In \bibinfo{booktitle}{\emph{Proceedings of the 17th Annual International Conference on Mobile Systems, Applications, and Services}}. \bibinfo{pages}{14--26}.
\newblock


\bibitem[Xu et~al\mbox{.}(2024)]%
        {mmear}
\bibfield{author}{\bibinfo{person}{Xiangyu Xu}, \bibinfo{person}{Yu Chen}, \bibinfo{person}{Zhen Ling}, \bibinfo{person}{Li Lu}, \bibinfo{person}{Junzhou Luo}, {and} \bibinfo{person}{Xinwen Fu}.} \bibinfo{year}{2024}\natexlab{}.
\newblock \showarticletitle{mmEar: Push the Limit of COTS mmWave Eavesdropping on Headphones}. In \bibinfo{booktitle}{\emph{IEEE INFOCOM 2024-IEEE Conference on Computer Communications}}. IEEE, \bibinfo{pages}{351--360}.
\newblock


\bibitem[Zhang et~al\mbox{.}(2023)]%
        {500noiserange2}
\bibfield{author}{\bibinfo{person}{Baoqing Zhang}, \bibinfo{person}{Yubin Rao}, \bibinfo{person}{Yunyi Guo}, {and} \bibinfo{person}{Wangqiang Xiao}.} \bibinfo{year}{2023}\natexlab{}.
\newblock \showarticletitle{Noise Reduction Performance of Metamaterials Sound Insulation Plate}. In \bibinfo{booktitle}{\emph{International Conference on Urban Climate, Sustainability and Urban Design}}. Springer, \bibinfo{pages}{656--666}.
\newblock


\bibitem[Zhang et~al\mbox{.}(2021a)]%
        {r4}
\bibfield{author}{\bibinfo{person}{Guoming Zhang}, \bibinfo{person}{Xiaoyu Ji}, \bibinfo{person}{Xinfeng Li}, \bibinfo{person}{Gang Qu}, {and} \bibinfo{person}{Wenyuan Xu}.} \bibinfo{year}{2021}\natexlab{a}.
\newblock \showarticletitle{EarArray: Defending against DolphinAttack via Acoustic Attenuation.}. In \bibinfo{booktitle}{\emph{NDSS}}.
\newblock


\bibitem[Zhang et~al\mbox{.}(2021b)]%
        {meta}
\bibfield{author}{\bibinfo{person}{Jin Zhang}, \bibinfo{person}{Wei Rui}, \bibinfo{person}{Chengrong Ma}, \bibinfo{person}{Ying Cheng}, \bibinfo{person}{Xiaojun Liu}, {and} \bibinfo{person}{Johan Christensen}.} \bibinfo{year}{2021}\natexlab{b}.
\newblock \showarticletitle{Remote whispering metamaterial for non-radiative transceiving of ultra-weak sound}.
\newblock \bibinfo{journal}{\emph{Nature Communications}} \bibinfo{volume}{12}, \bibinfo{number}{1} (\bibinfo{year}{2021}), \bibinfo{pages}{3670}.
\newblock


\bibitem[Zhang et~al\mbox{.}(2015)]%
        {AccelWord}
\bibfield{author}{\bibinfo{person}{Li Zhang}, \bibinfo{person}{Parth~H Pathak}, \bibinfo{person}{Muchen Wu}, \bibinfo{person}{Yixin Zhao}, {and} \bibinfo{person}{Prasant Mohapatra}.} \bibinfo{year}{2015}\natexlab{}.
\newblock \showarticletitle{Accelword: Energy efficient hotword detection through accelerometer}. In \bibinfo{booktitle}{\emph{Proceedings of the 13th Annual International Conference on Mobile Systems, Applications, and Services}}. \bibinfo{pages}{301--315}.
\newblock


\bibitem[Zhu et~al\mbox{.}(2023)]%
        {r14}
\bibfield{author}{\bibinfo{person}{Yihuan Zhu}, \bibinfo{person}{Ruizhi Dong}, \bibinfo{person}{Dongxing Mao}, \bibinfo{person}{Xu Wang}, {and} \bibinfo{person}{Yong Li}.} \bibinfo{year}{2023}\natexlab{}.
\newblock \showarticletitle{Nonlocal Ventilating Metasurfaces}.
\newblock \bibinfo{journal}{\emph{Physical Review Applied}} \bibinfo{volume}{19}, \bibinfo{number}{1} (\bibinfo{year}{2023}), \bibinfo{pages}{014067}.
\newblock


\end{thebibliography}
\end{document}